\documentclass[preprint,12pt]{larticle}
\usepackage[a4paper, total={6.18in, 10in}]{geometry}
\usepackage[utf8]{inputenc}

\usepackage{amssymb}
\usepackage{amsthm}
\usepackage{amsmath}

\usepackage{tikz}
\usepackage{hhline}
\usepackage{soul}

\usepackage{mathtools}

\usepackage{array}
\newcolumntype{P}[1]{>{\centering\arraybackslash}p{#1}}

\usepackage[symbol]{footmisc}

\newcommand{\abs}[1]{\left\vert#1\right\vert}

\begin{document}

\begin{frontmatter}

\title{Effects of network topology and trait distribution on collective decision making}

\author{Pengyu Liu\footnotesize{$^{1,2}$}\footnote{To whom correspondence should be addressed; e-mail: penliu@ucdavis.edu.}}

\author{Jie Jian\footnotesize{$^3$}}

\address{\footnotesize{$^1$}Department of Microbiology and Molecular Genetics, University of California, Davis, Davis, CA 95616, USA}
\address{\footnotesize{$^2$}Department of Mathematics, Simon Fraser University, Burnaby, BC V5A 1S6, Canada}
\address{\footnotesize{$^3$}Department of Statistics and Actuarial Science, University of Waterloo, Waterloo, ON N2L 3G1, Canada}

\begin{significance}
Individual-level interactions shape societal or economic processes such as infectious diseases spreading, stock prices fluctuating and public opinion shifting. Understanding how different individuals interacting affect collective outcomes is more important than ever, as the internet and social media develop. Social networks representing individuals' influence relations and distributions of individuals with different traits determining decision making play key roles  in understanding the connections between individual-level interactions and societal or economic outcomes. In this work, we develop novel mathematical models to analyze the effects of network topology and trait distribution on collective decision making. Our findings suggest that unstable collective decisions are more probable when individuals are more connected, networks are centralized or individuals with different traits are excessively clustered.

\end{significance}

\begin{abstract}
Social networks play an important role in analyzing the impact of individual-level interactions on societal or economic outcomes. 
We model interactive decision making for a community of individuals with different traits, represented by a social network with trait-attributed nodes.
We develop a deterministic process generating a sequence of choices for each individual based on a trait-attributed social network, initial choices of individuals and a set of predetermined trait-dependent rules for making decisions.
The object of interest is the sequence of cumulative sum of choices over all individuals, which we call the cumulative sequence and consider as an index of collective decisions.
We observe that, in a time period, a cumulative sequence can be unpredictable or predictable showing a repeated pattern either escalating to an extreme or constantly oscillating.
We consider that predictable cumulative sequences represent unstable collective decisions of communities either extremizing or internally conflicting, while unpredictable cumulative sequences show stable changes.
We analyze the effects of network topology and trait distribution on the probability of cumulative sequences being predictable, escalating and oscillating by simulations.
Our findings include that unstable collective decisions are more probable as network density increases, that centralized networks are more likely to have unstable collective decisions and that networks with excessively clustered or scattered conformists and rebels tend to produce unstable cumulative sequences.
We discuss the potential of the model as a framework for studying individuals with different traits on a social network directly and indirectly interacting in decision making.
\end{abstract}

\end{frontmatter}

\section{Introduction}
\label{S:1}

An individual in a community receives information when making a decision or a choice.
The information includes suggestions, preferences and decisions of trusted individuals and indices such as stock prices reflecting collective decisions of a community. 
Direct social interactions can happen with the information from trusted individuals, and indirect social interaction can occur with the information reflecting collective decisions.
Individuals buying or selling stocks based on recommendations from trusted individuals is an example of direct social interaction.
Meanwhile, people trading their shares determines stock prices, and individuals exchanging stocks based on the prices is an example of indirect social interaction. 
Indeed, what information to use and how to process the information to make a decision depend on an individual's trait.
For example, there are individuals choosing to listen to music liked by their friends, and there also exist individuals who want to be distinctive and reject popular music, which is defined by collective choices of a community.
In this work, we introduce a deterministic framework for modeling individuals with different traits directly and indirectly interacting in decision making.

Studying the dynamics of social interaction in decision making has stimulated the development of various mathematical models in different fields.
The discrete choice models explain and forecast an individual's choice from a set of alternatives \cite{Anderson1992}, for example, which college to attend \cite{Fuller1982} and which vehicle to purchase \cite{Train2007}.
Classic discrete choice models assume that individuals make decisions for utility maximization, and non-rational behavior in decision making is modeled by random utility \cite{Manski1977}. 
In \cite{Brock2001}, the authors extended discrete choice models to analyze decision making with social interactions by integrating factors reflecting the inclination of an individual to conform to the choice of others.
In \cite{Cont2010}, the authors developed a characteristic-stratified discrete choice model to understand how interaction of individuals from different peer groups affects equilibrium behavior in decision making.
The voter model \cite{Holley1975} and its generalization \cite{Castellano2009} are simple stochastic processes describing opinion formation or decision making, which are also applicable to study phase transition in statistical physics and to model the dynamics of language death \cite{Abrams2003}. 
Compared with discrete choice models, an individual in voter models does not make a decision for utility maximization but takes the majority choice of a set of random neighbors on the individual's social network with a probability, so the voter models describe direct social interaction by nature.
In \cite{Kruege2017}, the authors generalized voter models to study the effects of conformity and anticonformity on polarization in opinion dynamics.
In social physics and social psychology, understanding opinion formation has stimulated models studying how individuals make decisions under pressures from others \cite{Latane1981,Nowak1990,Sznajd2000} and models analyzing how proportions of individuals with different traits affect the equilibrium behavior in interactive decision making \cite{Cheon2016,Galam2005}

We develop a mathematical model of a community of individuals with different traits interacting in decision making.
The model is individual-based \cite{Bonabeau2002}.
We model a community of individuals and influence relations by a social network, where nodes represent individuals, and edges indicate two individuals influencing each other in decision making.
Each individual in a community has a trait, and the trait determines how the individual makes a decision with the received information.
Here, we only focus on two traits: being a conformist and being an anticonformist or rebel.
The model is a discrete-time deterministic process.
At each time point, every individual makes a choice of either $-1$ or $1$ simultaneously following a set of predetermined trait-dependent rules without random factors.
This process produces a time series or sequence of choices for each individual, and it being deterministic allows us to study these time series, their sum and accumulation in addition to the equilibrium behavior. 
The time series of sum and cumulative sum of choices over all individuals in a community are indices reflecting collective decisions of the community, representing, for example, stock prices, the community's position on the left-right political spectrum or the popularity of mainstream music. 
As discussed in \cite{Granovetter2005}, the topology of a social network can affect social or economic outcomes.
In this work, we study the effects of network topology and trait distribution on the sequences of sum and cumulative sum of choices.
We control the network topology by developing a generalized Erd\"os-R\'enyi model to generate random networks with three parameters regulating the size, the density and the heterogeneity of generated random networks, and we control the trait distribution on a network by developing a trait attributing random process with two parameters regulating the proportion of individuals with each trait and the extent of mixing for individuals with different traits.
We also use a real social network available in \cite{SnapNets,Rozemberczki2021} to study the effects of trait distribution on the sequence of collective decisions.
Lastly, we discuss the potential for this model as a framework of studying individuals with different traits on a social network directly and indirectly interacting in decision making.

\section{Methods}
\label{S:2}

\subsection{Model assumptions and basic definitions}
\label{S:2.1}

We consider a community of $n$ individuals.
We assume that each individual makes a {\em sequence of choices} from the binary state space $\Omega = \{-1,1\}$ and that individuals' sequences of choices have the same length of $t+1$ steps. 
So, the model is discrete-time, and we use ``sequence'' and ``time series'' interchangeably.
We use an integer $1\leq i\leq n$ to represent an individual in the community and denote the sequence of choices made by individual $i$ as $C(i,\cdot) = [c(i,0),c(i,1),...,c(i,t)]$.
The element $c(i,k)$ in the sequence, being either $-1$ or $1$, is the {\em choice} of individual $i$ at step $k$.
The choices of all individuals in the community at step $k$ also form a sequence $C(\cdot,k) = [c(1,k),c(2,k),...,c(n,k)]$, which we call the {\em choice pattern} of the community at step $k$. 
In particular, the choice pattern at step zero $C(\cdot,0)$ is the {\em initial choices} of the community.
The time series of choice patterns  $C=[C(\cdot,0),C(\cdot,1),...,C(\cdot,t)]$ is the community's {\em sequence of choice patterns}.

We say that two individuals in a community are {\em related} if they influence each other in decision making.
We assume that influence in decision making is mutual, that there is no self influence and that influence is indifferent with the same strength.
With respect to these assumptions, we model a community of individuals and how they are related in decision making by a {\em social network} $N$ which is undirected and without loops or multiple edges.
Each node in a social network represents an individual, so we use ``node'' and ``individual'' interchangeably.
An edge connecting two nodes in a social network indicates a pair of related individuals.
Two related individuals are {\em neighbors} of each other on a social network.

Every individual in a community has a {\em trait} of either being a {\em conformist} or being a {\em rebel} in decision making.
The trait affects how an individual makes choices.
In a community of $n$ individuals, the function $f:\{1,2,..,n\}\to\{\text{conformist},\text{ rebel\}}$ assigning a trait to each individual defines a {\em trait distribution}.
We assume that every individual makes a choice simultaneously at every step and that an individual's choice at step $k$ is determined by the choices of the individual's neighbors at step $k-1$. 
We list explicit rules for an individual to make a choice at each step in Table~\ref{t1}. 

\begin{table}[ht]
    \centering
    \begin{tabular}{|c||c|c|}
    \hline
         Trait & Step $k-1$ & Step $k$ \\ \hhline{|=||=|=|}
         Conformist & More neighbors who chose $-1$ & $-1$ \\ \cline{2-3}
                               & More neighbors who chose $1$ & $1$ \\ \cline{2-3}
                               & Equal numbers of neighbors who chose $-1$ and $1$ & $c(i,k-1)$ \\ \hline
         Rebel      & More neighbors who chose $-1$ & $1$ \\ \cline{2-3}
                               & More neighbors who chose $1$ & $-1$ \\ \cline{2-3}
                               & Equal numbers of neighbors who chose $-1$ and $1$ & $c(i,k-1)$  \\ \hline
    \end{tabular}
    \caption{{\bf Rules for individual $i$ to make a choice at step $k$.}}
    \label{t1}
\end{table}

Note that the rules are deterministic, so with a given social network representing how a community of individuals influence each other in decision making, the distribution of traits on the social network and the initial choices of the community, each individual's sequence of choices is determined. 
Consider a community of $n$ individuals.
The {\em collective choice} of the community at step $k$ is the sum of choices over all individuals at step $k$ and denoted by $s(k) = \sum_{l=1}^n c(l,k)$.
The time series $s = [s(0),s(1),s(2),...,s(t)]$ is the {\em collective sequence} of the community's choices.
Analogously, the {\em cumulative choice} of the community at step $k$ is the cumulative sum of choices over all individuals at step $k$ and denoted by $S(k) = \sum_{l=0}^k s(l)$, and the time series $S = [S(0),S(1),S(2),...,S(t)]$ is the {\em cumulative sequence} of the community's choices. 

\subsection{Toric lattices and random networks}
\label{S:2.2}

We use two classes of network topologies as models for the influence relations of a community of individuals.
For simplicity and visualizing trait distribution, the first class is lattices with no boundary. 
A {\em toric lattice} of size $m$ is constructed such that there exists a node at every integer coordinate $(x,y)$ in the plane for integers $0 \leq x,y \leq m-1$ and no node at other coordinates, so the toric lattice has $m^2$ nodes. 
Each node in a toric lattice is only related to the eight surrounding nodes, and to make the lattice boundaryless, the nodes on a boundary of a lattice are related to some nodes on the opposite boundary as in the example displayed in Supplementary Figure~\ref{sf0}. See supplementary material for more detail.

We develop a generalized Erd\"os-R\'enyi model to generate random networks as another class of network topology.
The {\em degree} of a node in a network is the number of edges incident to it.
We are interested in the mean value and the standard deviation of the degrees of nodes in a network, and we call the two quantities the {\em mean degree} and the {\em degree deviation} of the network respectively.
Without ambiguity, we denote the mean degree of a social network by $\mu$ and the network's degree deviation by $\sigma$. 
The mean degree of a social network, proportional to the network density \cite{Granovetter2005}, reflects the level of connectedness of individuals in a community, and the degree deviation of a social network is known to represent network heterogeneity \cite{Snijders1981}.
The generalized Erd\"os-R\'enyi model can generate random networks with a certain number of nodes, a specific mean degree and a degree deviation regulated by a parameter. The model allows us to study the effects of network size (number of nodes), network density (mean degree) and network heterogeneity (degree deviation) on the collective and cumulative sequences of a community's choices with control.

To generate a random network with $n$ nodes and mean degree $\mu$, we add $n\mu/2$ edges successively to pairs of nodes in the network as follows.
Let $\eta$ be the {\em heterogeneity parameter} regulating the degree deviation of a network, and $d(i)$ be the degree of the node representing individual $i$, which we call node $i$ for short.
To select the two end nodes of an edge to be added, we assign each node a weight which determines the probability of the node being selected.
To select the first end node, we assign a weight $w(i)=(1+d(i))^{\eta}$ to node $i$ when $d(i)\leq n-1$ and $w(i)=0$ to the node when $d(i) > n-1$ to avoid multiple edges, then the probability of node $i$ being selected as the first end node is $p(i)=w(i)/\sum_{l=1}^n w(l)$.
Suppose that node $i$ is selected as the first end node.
To select the second end node, we assign node $i$ a weight $v(i)= 0$ to avoid loops and $v(j) = 0$ to node $j$ when node $j$ is a neighbor of node $i$ or $d(j) > n-1$ to avoid multiple edges; otherwise, we assign node $j$ a weight $v(j)=(1+d(j))^{\eta}$.
Similarly, the probability of node $j$ being selected as the second end node is $q(j)=v(j)/\sum_{l=1}^n v(l)$.

Note that when $\eta=0$, each possible edge of the network to be generated has the same probability to be added, so the model generates Erd\"os-R\'enyi random networks with $n$ nodes and $n\mu/2$ edges.
When $\eta < 0$, nodes with a high degree are less likely to be selected as an end node, so the generated networks are more regular with low degree deviation.
When $\eta > 0$, nodes with a high degree are more likely to be selected as an end node, so the generated networks are more centralized or star-like with high degree deviation.
See Supplementary Figure~\ref{sf1} for the relations between the heterogeneity parameter and degree deviations of generated random networks.

\subsection{Trait distribution}
\label{S:2.3}

Let $N$ be a social network representing a community of individuals and their relations in decision making. 
We attribute the traits of being a conformist and being a rebel to the nodes in $N$ and call the resulting network an {\em attributed network} denoted by $N'$.
We characterize trait distributions with two quantities: the number of rebels $r$ in a community and a parameter measuring the extent of mixing for individuals with different traits which is defined as the average number of conformist neighbors over all rebels.
We call the second quantity of an attributed network the {\em mixing parameter} and denote it by $\chi$.

We develop the following process to attribute traits to nodes of a social network so that the mixing parameter can vary in a wide range. 
Consider a network $N$ with $n$ nodes.
To attribute $r$ rebels and $n-r$ conformists to the nodes of $N$, we initially attribute all nodes in $N$ as conformists and then successively select $r$ nodes to be rebels with an {\em attributing parameter} $\alpha$ regulating the mixing parameter $\chi$.
To select $r$ rebels, we assign each node in the network weights which determine the probability of the node being selected.
To select the $k$-th rebel, we assign a weight $u(i,k) = 0$ when node $i$ is a rebel and a wight $u(i,k) = \alpha^m$ if node $i$ has $m$ rebel neighbors, then the probability for node $i$ to be selected as the $k$-th rebel is $p(i,k) = u(i,k)/\sum_{l=1}^n u(l,k)$.

Note that if the attributing parameter $\alpha = 1$, then the rebels are uniformly selected at random.
If $\alpha > 1$, the rebels are clustered and the mixing parameter is low.
For $0 < \alpha < 1$, the mixing parameter is high and the rebels are scattered.
See Supplementary Figure~\ref{sf2} for examples of attributed toric lattices with scattered and clustered rebels and relations between the attributing parameter $\alpha$ and the mixing parameter $\chi$.

\subsection{Predictability of cumulative sequences}
\label{S:2.4}

Mathematical proof shows that every collective sequence of choices eventually enters a unique period determined by the topology and the trait distribution of the attributed network and initial choices of the community, hence every cumulative sequence eventually shows a unique repeated pattern. See supplementary material for detailed arguments. 
Without ambiguity, we call both the unique period of the collective sequence and the repeated pattern of the corresponding cumulative sequence the {\em eventual period} of the sequences.
We define the subsequence of a cumulative sequence before its first eventual period the {\em pre-period subsequence}.
We denote the eventual period of a cumulative sequence by $P$ and its pre-period subsequence by $Q$.
The {\em length} of an eventual period $P$ is the number of steps that it spans, which is denoted by $L(P)$.
Similarly, the length of a pre-period subsequence denoted by $L(Q)$ is the number of steps the subsequence spans. 
The change in the cumulative sequence over the period $P$ is the {\em period gain} denoted by $\Delta P$.
We define the {\em gradient} of the eventual period by $\nabla P = \abs{\Delta P}/L(P)$, which we use to describe the asymptotic behavior of cumulative sequences. 
See supplementary material for examples and more detail about eventual periods.

We say that the cumulative (collective) sequence of a community's choices is {\em predictable} if there exists at least one complete eventual period in the first $t+1-\tau$ steps of the sequence. 
We consider the first $t+1-\tau$ steps instead of the total $t+1$ steps of the sequence because if there exists only one complete eventual period that ends close to step $t+1$, then it is hard to determine if the sequence contains the unique period.
For experiments in this paper, we set $t=10000$ and $\tau = 50$ unless otherwise stated.
If the cumulative (collective) sequence of a community's choices is not predictable, then it is {\em unpredictable}. 
To efficiently determine if a cumulative (collective) sequence is predictable without recording and comparing choice patterns, we develop a heuristic method. 
The heuristic method can determine predictability with an average accuracy over $99.4\%$.
See supplementary material for more detail about the heuristic method and its accuracy.

\subsection{Summary of parameters and experiments}
\label{S:2.5}

There are three factors regulating the deterministic process: the network topology, the trait distribution and the initial choices of a community of individuals. With the three factors pre-specified, the model generates a unique cumulative sequence of the community's choices
In Table~\ref{t2}, we summarize the parameters controlling the three factors and their values used in experiments for examining the effects of the three factors on the probability of cumulative sequences being predictable. 
Each parameter has a default value in the experiments: $n = 100$, $\mu = 8$, $\eta = 0$, $r = 50\%n$, $\alpha = 0.8$ and initial choices being $-1$ for all individuals. 
In experiments analyzing the effects of the number of nodes, $n$ takes 100 data points from 2 to 200 in increments of $2$.
In experiments studying the effects of the mean degree, $\mu$ takes 101 data points from 0 to 50 in increments of $0.5$.
In experiments examining the effects of the heterogeneity parameter, $\eta$ takes 101 data points from $-80$ to $20$ in increments of $1$ and 101 data points from $-2$ to $8$ in increments of $0.1$.
In experiments analyzing the effects of the number of rebels, $r$ takes 101 data points from $0\%n$ to $100\%n$ in increments of $1\%n$ and we round $r$ to the smaller integer if $r$ is not an integer.
In experiments studying the effects of the attributing parameter, $\alpha$ takes 79 data points from $0.05$ to $2$ in increments of $0.025$.
For each data point in these experiments, we generate 10000 random networks with other parameters taking default values and compute the proportion of predictable cumulative sequences as the probability of cumulative sequences being predictable.
In experiments analyzing the effects of initial choices, we randomly generate initial choices such that each individual has a probability of $0.5$ to choose $-1$ or $1$. 
We vary the network topology parameters and the trait distribution parameters independently in their ranges and use the data points for the parameters as described above.
We generate 100 attributed random networks for each data point, and for each random network, we generate cumulative sequences with 100 random initial choices.
For each attributed random network, we compute the proportion of initial choices that produce cumulative sequences with the majority predictability.
We scale the ranges for each parameter listed in Table~\ref{t2} to the same range of $[0,1]$ linearly for comparison.

\begin{table}[ht]
    \centering
    \begin{tabular}{|P{100px}|P{75px}||P{57px}|P{57px}|P{57px}|}
        \hline
        Factor & Parameter & \multicolumn{3}{c|}{Experiments}\\ \cline{3-5}
                           &  & Network topology & Trait distribution & Initial choices \\
        \hhline{|=|=||=|=|=|}
        Network Topology & Number of nodes $(n)$ & $[2,200]$ & $100$ & $100$ \\ \cline{2-5}
         & Mean degree $(\mu)$ & $[0,50]$ & $8$ & $8$ \\ \cline{2-5}
         & Heterogeneity parameter $(\eta)$ & $[-80,20]$ & $0$ & $0$ \\ \hhline{|=|=||=|=|=|}
         Trait distribution & Number of rebels $(r)$ & $50\%n$ & $[0,100]$ & $50$ \\ \cline{2-5}
         & Attributing parameter $(\alpha)$ & $0.8$ & $(0,2]$ & $0.8$ \\ \hhline{|=|=||=|=|=|}
         Initial choices & Initial choices & $-1$ for all &  $-1$ for all & Uniformly at random \\ \hline
    \end{tabular}
    \caption{{\bf Summary of parameters and their values used in experiments.}}
    \label{t2}
\end{table}

In addition, we use a Twitch user-user network of gamers who stream in Portuguese (PT) as a real social network topology for our study \cite{SnapNets,Rozemberczki2021}. 
The network has 1912 nodes, with a mean degree of 32.74 and a degree deviation of 55.85.
We attribute traits of being a conformist and being a rebel to the nodes of the real social network and study the effects of the trait distribution on the probability of cumulative sequences being predictable. 
In the experiments, the number of rebels takes 192 data points varying from $0\%n$ to $100\%n$ and the attributing parameter takes 192 data points varying from $0.05$ to $2$.
For each data point, we generate 100 trait distributions and compute the proportion of predictable cumulative sequences as the probability of cumulative sequences being predictable.

\section{Results}
\label{S:3}

\subsection{Cumulative sequences}
\label{S:3.1}

\begin{figure}[p]
\begin{center}
\includegraphics[scale=0.5]{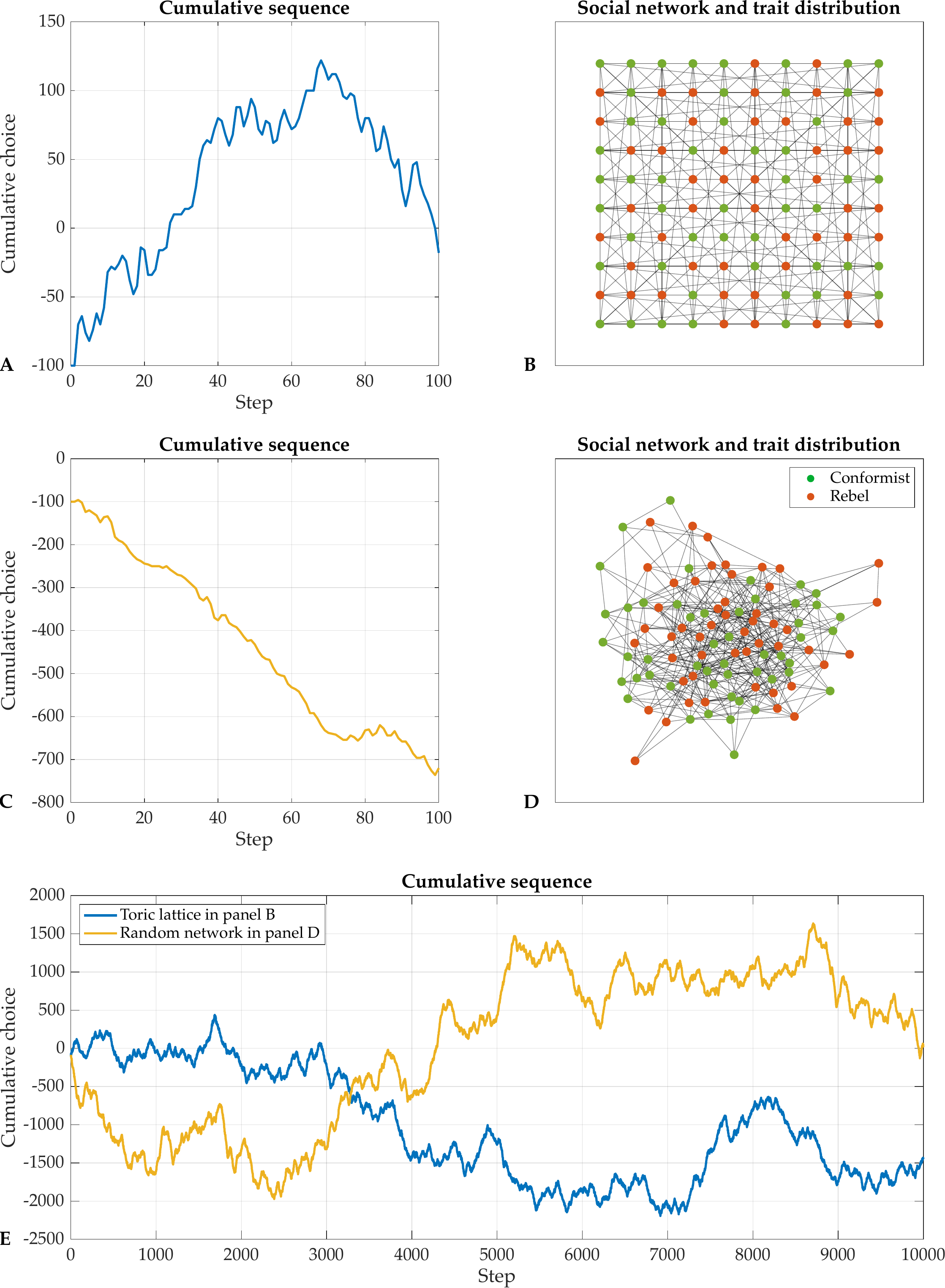}
\renewcommand{\figurename}{Figure}
\caption{{\bf Simple deterministic processes generate unpredictable cumulative sequences.} The first 100 steps of the cumulative sequence of choices (panel A) of the community represented by the attributed toric lattice of size $m=10$ generated with $r =50$ and $\alpha = 0.7$ (panel B), and the first 100 steps of the cumulative sequence of choices (panel C) of the community represented by the attributed random network generated with $n=100$, $\mu =8$, $\eta = 0$, $r= 50$ and $\alpha = 0.9$ (panel D).
The first 10000 steps of the cumulative sequences (panel E) are displayed in panels A and C. 
The initial choices for both attributed networks are $-1$ for all individuals. }\label{f2}
\end{center}
\end{figure}

We are mainly interested in the cumulative sequence of a community's choices and how the network topology and the trait distribution of the community affect the cumulative sequence.  
We consider the cumulative sequences as indices reflecting the changes of a community's collective decisions or opinions on a matter over time, for example, stock prices or a community moving left or right on the political spectrum.
We observe that cumulative sequences determined by the rules described in Table~\ref{t1} and the three factors listed in Table~\ref{t2} can be unpredictable and resemble random walks. See Figure~\ref{f2}.
Moreover, we can not predict the future movements of the cumulative sequences with the subsequences from past steps.
For instance, the trends observed in the first 100 steps of cumulative sequences displayed in panels A and C of Figure~\ref{f2} do not suggest the trends in the 10000-step sequences displayed in panel E. 
As argued in Section~\ref{S:2.4} and supplementary material, if we do not terminate the process, every cumulative sequence will enter its unique eventual period determined by the three factors listed in Table~\ref{t2}.
So, a cumulative sequence consists of two parts: the pre-period subsequence which can have a length of zero and the eventual period which may not be complete. 
Supplementary Figure~\ref{sf3} displays examples of pre-periods subsequences and eventual periods.
The observed unpredictable cumulative sequences in the first $t=10000$ steps can be part of the pre-period subsequence, part of the eventual period or a mixture of the end of the pre-period subsequence and the beginning of the eventual period.

For homogeneous attributed networks displayed in Supplementary Figure~\ref{sf4}, we can deduce their cumulative sequences. 
Random networks of all conformists and toric lattices with homogeneously clustered conformists and rebels generate predictable cumulative sequences escalating to an extreme with $\nabla P > 1$.
In contrast, random networks of all rebels and toric lattices with homogeneously mixed conformists and rebels generate predictable cumulative sequences oscillating with $\nabla P = 0$.
See supplementary material for more detail. 
Suppose that the cumulative sequences represent the changes of communities' positions on the left-right political spectrum.
The escalating predictable cumulative sequences with $\nabla P > 1$ indicate the communities are fast extremizing, and the oscillating predictable cumulative sequences with $\nabla P = 0$ suggest the communities have constant internal conflicts without any movement.
Neither the escalating cumulative sequences nor the oscillating ones represent stable movements of communities on the political spectrum.
The unpredictable cumulative sequences, however, show movements without extremizing or constant internal conflicts.
In the following sections, we demonstrate the effects of the three factors listed in Table~\ref{t2} on the probabilities of cumulative sequences being predictable, escalating ($\nabla P > 1$) and oscillating ($\nabla P = 0$), which we call the {\em probability of predictable sequences}, the {\em probability of escalating sequences} and the {\em probability of oscillating sequences} respectively for short.

\subsection{The effects of network topology}
\label{S:3.2}

\begin{figure}[h!p]
\begin{center}
\includegraphics[scale=0.5]{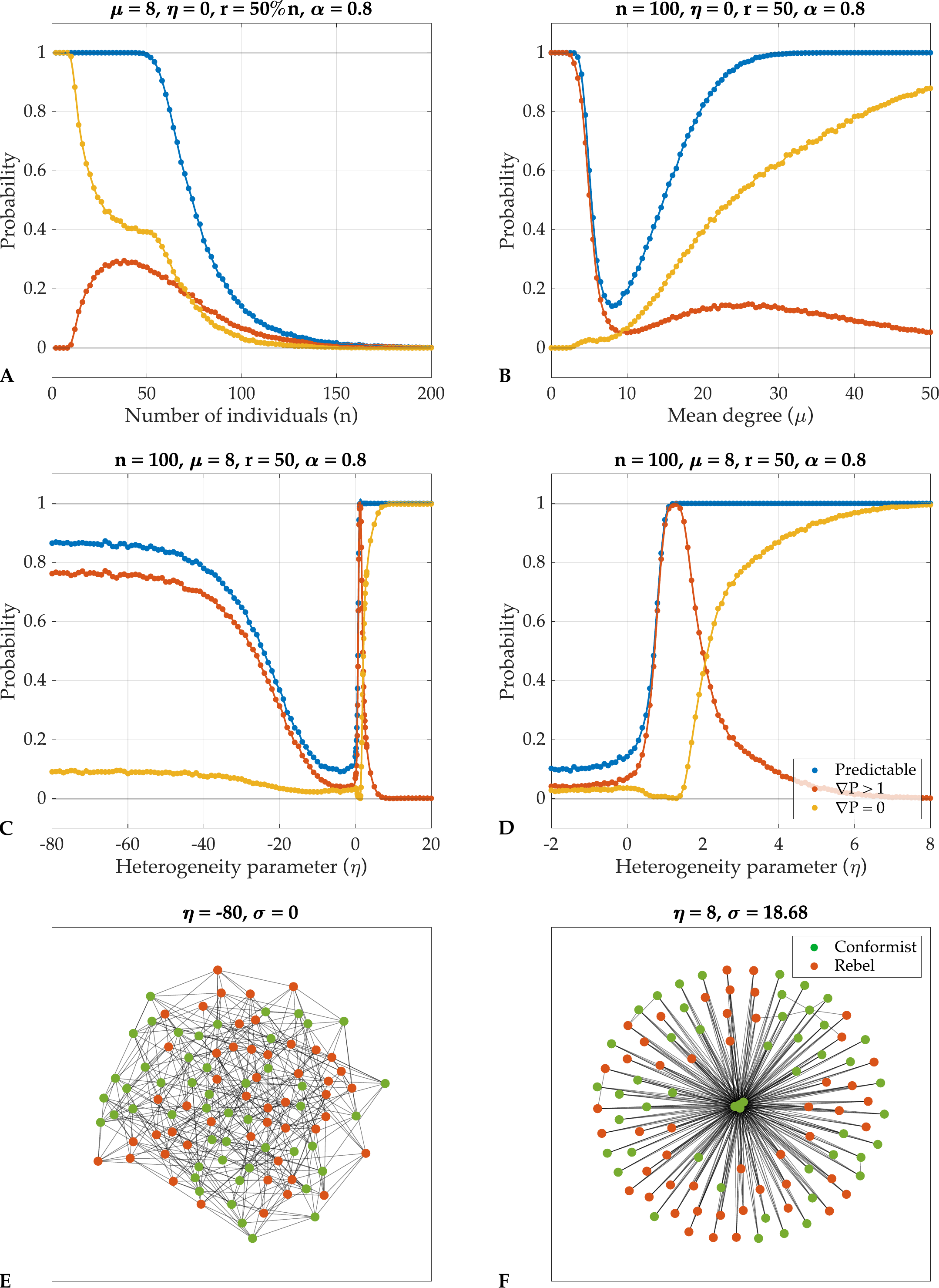}
\renewcommand{\figurename}{Figure}
\caption{{\bf Effects of the three network topology parameters on the probability of predictable sequences.} The relations between the probabilities of predictable, escalating and oscillating sequences and the number of individuals (nodes) $n$ (panel A), the mean degree $\mu$ (panel B) and the heterogeneity parameter $\eta$ from $-80$ to $20$ (panel C) and from $-2$ to $8$ (panel D); each data point is computed with 10000 random networks and their corresponding cumulative sequences with initial choices $-1$ for all individuals; smooth fitted curves are added for visualization.
A regular random network generated with $n = 100$, $\mu = 8$, $\eta = -80$, $r =50$ and $\alpha = 0.8$ (panel E).
A star-like random network generated with $n = 100$, $\mu = 8$, $\eta = 8$, $r =50$ and $\alpha = 0.8$ (panel F).}\label{f3}
\end{center}
\end{figure}

The relation between the number of individuals in a community and the probability of predictable displayed in Panel A of Figure~\ref{f3} shows that smaller communities have a higher probability of predictable sequences and that the probability decreases as the number of individuals increases.
This suggests that smaller communities are more likely to extremize or internally conflict, while larger communities are more probable to have stable unpredictable collective decisions, even though all individuals are non-rational conformists and rebels.
Smaller communities with more rebels are prone to internal conflicts, while smaller communities with more conformists are more likely to extremize; see Supplementary Figure~\ref{sf5}.

According to the rules listed in Table~\ref{t1}, an individual who is not related to any other individual in decision making chooses the initial choice at every step.
Communities with a low mean degree (density) have many such individuals, and trivially, the predictable cumulative sequences are more likely to be escalating.
The relation between the mean degree and the probability of predictable sequences displayed in panel B of Figure~\ref{f3} indicates that the probability of predictable cumulative sequences is also high for communities with a high mean degree. Supplementary Figure~\ref{sf6} suggests that high-density communities with more rebels are prone to internal conflicts, while high-density communities with more conformists are more likely to extremize.
Moreover, high-density communities with scattered rebels are more probable to internal conflict.
These results provide an explanation for more frequent conflicts and the increasing number of extremizing communities nowadays, as more individuals are being connected by various internet social media and influencing each other in decision making.

The heterogeneity parameter regulates the degree deviation of generated networks; see Supplementary Figure~\ref{sf1} for relations between the two quantities.
When $\eta = -80$, the generated network is regular (every node in the network has the same degree and $\sigma = 0$) with a near-one probability, and when $\eta = 8$, the generated network is centralized or star-like; see panel E and F in Figure~\ref{f3} for examples of regular and star-like networks.
Regular networks have high probabilities of predictable sequences, which depends on other parameters; see panel C in Figure~\ref{f3} and Supplementary Figure~\ref{sf7}. 
Specifically, the probability of predictable sequences are relatively low for regular networks with more individuals ($n = 150$), a higher density ($\mu = 12$) or more scattered rebels ($\alpha = 0.6$).
Moreover, the predictable cumulative sequences are more likely to be escalating for regular networks unless more than $80\%$ of the individuals are rebels; see panel C in Figure~\ref{f4}.
Star-like networks have near-one probabilities of predictable sequences regardless of other parameters; see panel D in Figure~\ref{f3} and Supplementary Figure~\ref{sf7}. 
If a star-like network has more rebels, then the predictable cumulative sequences are escalating, and if a star-like network has more conformists, then the predictable cumulative sequences are oscillating; see panel E in Figure~\ref{f4}.
These results indicate that communities of only non-rational conformists and rebels with star-like network topologies are prone to extremizing or internal conflicts, and the random networks generated with $\eta$ slightly below zero having the lowest probabilities of predictable sequences hints the benefit of network (e.g. internet) decentralization. 

\subsection{The effects of trait distribution}
\label{S:3.3}

\begin{figure}[h!p]
\begin{center}
\includegraphics[scale=0.5]{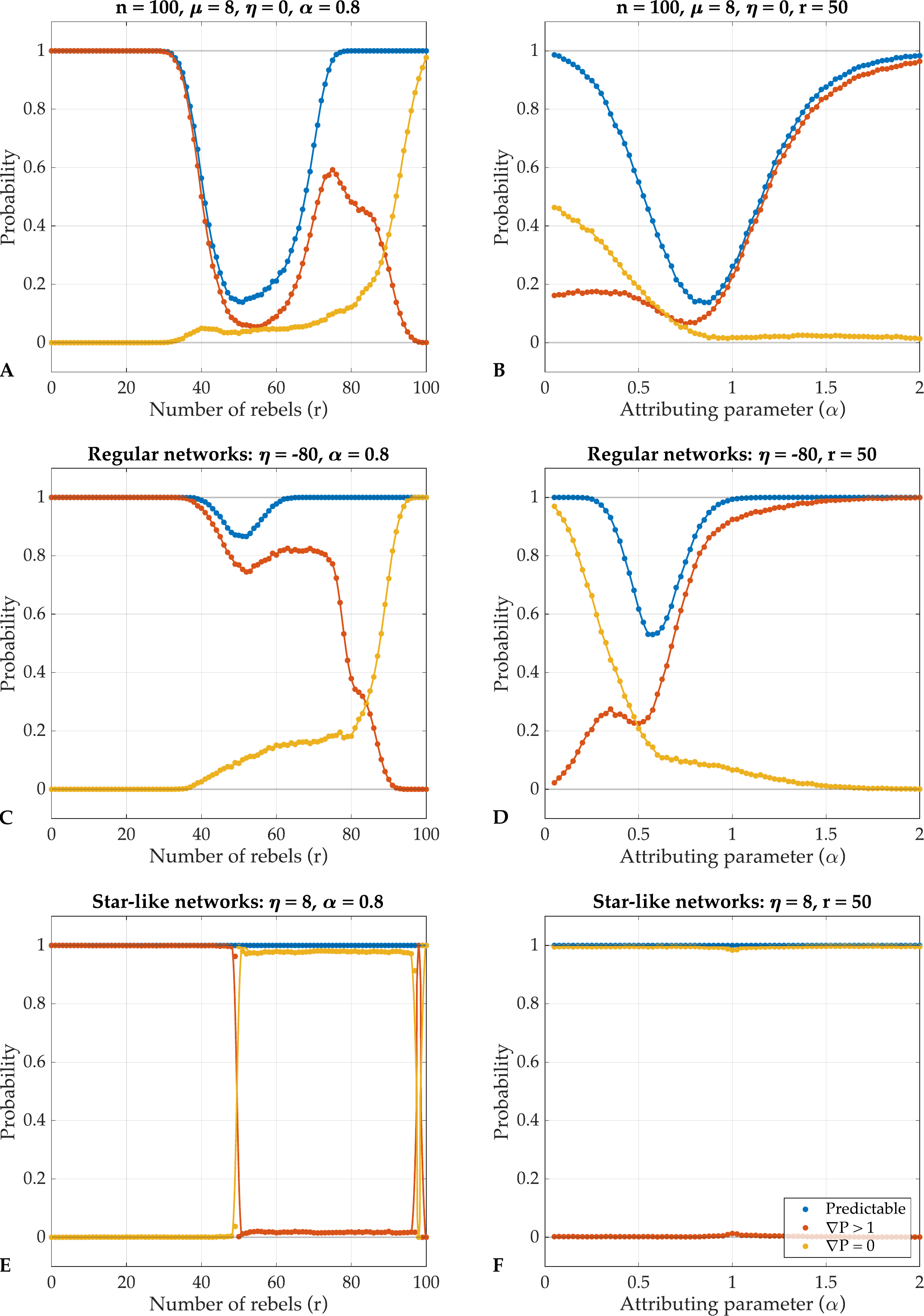}
\renewcommand{\figurename}{Figure}
\caption{{\bf Effects of the two trait distribution parameters on the probability of predictable sequences.} The relations between the probability of predictable, escalating and oscillating sequences and the number of rebels $r$ for networks with $n = 100$, $\mu = 8$, $\eta = 0$ and $\alpha = 0.8$ (panel A), regular networks with $\eta = -80$ (panel C) and star-like networks with $\eta = 8$ (panel E), and the relations between the probabilities and the attributing parameter $\alpha$ for networks with $n = 100$, $\mu = 8$, $\eta = 0$ and $r = 50$ (panel B) , for regular networks with $\eta = -80$ (panel D) and star-like networks with $\eta = 8$ (panel F). Each data point is computed with 10000 random networks and their corresponding cumulative sequences with initial choices $-1$ for all individuals. Smooth fitted curves are added for visualization.}\label{f4}
\end{center}
\end{figure}

Figure~\ref{f4} and Supplementary Figure~\ref{sf8} show that communities with fewer than $20\%$ of the individuals being rebels have near-one probabilities of predictable sequences, and the cumulative sequences are escalating.
Communities with more than $80\%$ of the individuals being rebels also have near-one probabilities of predictable sequences, and in most of the settings, more predictable cumulative sequences are oscillating when more than $90\%$ of the individuals are rebels.

The attributing parameter controls how conformists and rebels mix on a social network.
The relations between the attributing parameter and the mixing parameter are displayed in Supplementary Figure~\ref{sf2}, where panels A and B show examples of attributed networks with excessively scattered and clustered individuals with different traits.
Figure~\ref{f4} and Supplementary Figure~\ref{sf9} show that communities with excessively clustered or scattered conformists and rebels have high probabilities of predictable sequences, and the predictable cumulative sequences of communities with excessively clustered conformists and rebels are more likely to escalate than to oscillate in all studied parameter settings except when the network topology is star-like. 
Star-like networks have constant mixing parameters as the attributing parameters varies as displayed in panel E of Supplementary Figure~\ref{sf2}, and the predictable cumulative sequences being oscillating for any attributing parameter displayed in panel F of Figure~\ref{f4} is due to the number of rebels.

These results suggest that communities with even proportions of conformists and rebels are more likely to have stable collective decisions.
In addition to the proportions of individuals with different traits, how conformists and rebels are mixed on the social network and interact also plays an important role in determining the probability of stable collective decisions. 
If a community has even proportions of conformists and rebels, but they are excessively clustered, then the community is also prone to extremizing.

\subsection{The effects of initial choices}
\label{S:3.4}

\begin{figure}[p]
\begin{center}
\includegraphics[scale=0.5]{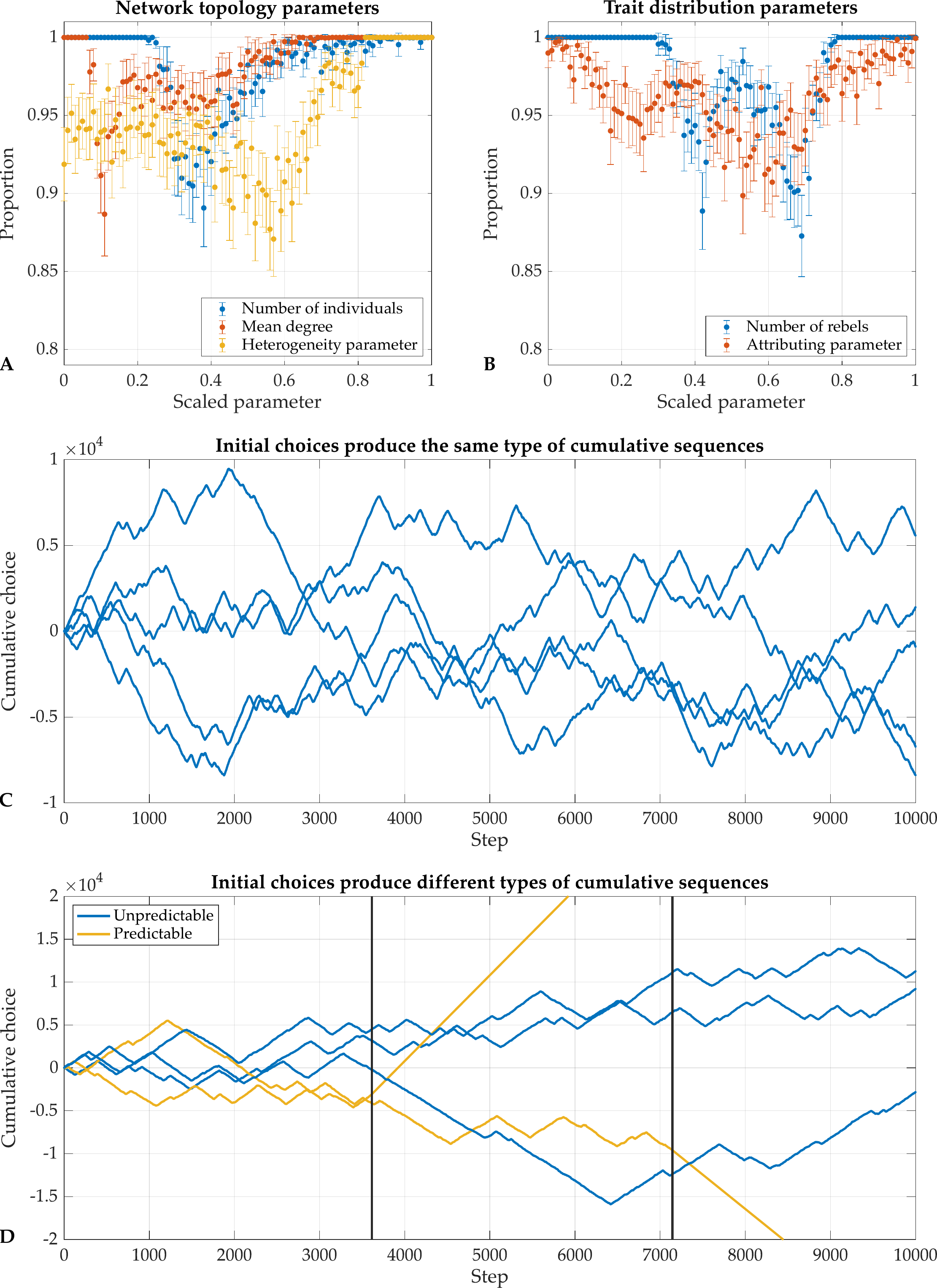}
\renewcommand{\figurename}{Figure}
\caption{{\bf Effects of initial choices on the probability of predictable sequences.} The average proportions of initial choices that generate cumulative sequences of majority predictability for varying network topology parameters (panel A) and trait distribution parameters (panel B). 
Each data point represents the mean proportion of 100 random initial choices over 100 random networks, and the error bars show $95\%$ confidence interval of the mean.
Five cumulative sequences of the same predictability produced by five random initial choices (panel C) and three unpredictable and two predictable cumulative sequences produced by five random initial choices (panel D) on two attributed networks generated with $n=100$, $\mu=8$, $\eta=0$, $r= 50$ and $\alpha = 1$ respectively; for predictable sequences, complete eventual periods are displayed between vertical lines, and pre-period subsequences are before the first vertical line.}\label{f5}
\end{center}
\end{figure}

Initial choices do not affect the predictability of cumulative sequences generated by an attributed network by much.
Panel A in Figure~\ref{f5} shows the proportions of random initial choices that generate cumulative sequences with majority predictability with different network topology parameters, and panel B shows the proportions for the trait distribution parameters.
On average, over $85\%$ initial choices generate cumulative sequences of the same predictability with an attributed network. 
However, the cumulative sequences are sensitive to initial choices.
The cumulative sequences generated with different initial choices displayed in panels C and D of Figure~\ref{f5} show different trajectories, though we can observe that the two sets of sequences have distinguishable characteristics determined by the attributed networks.

\subsection{Real social network}
\label{S:3.5}

\begin{figure}[h!p]
\begin{center}
\includegraphics[scale=0.5]{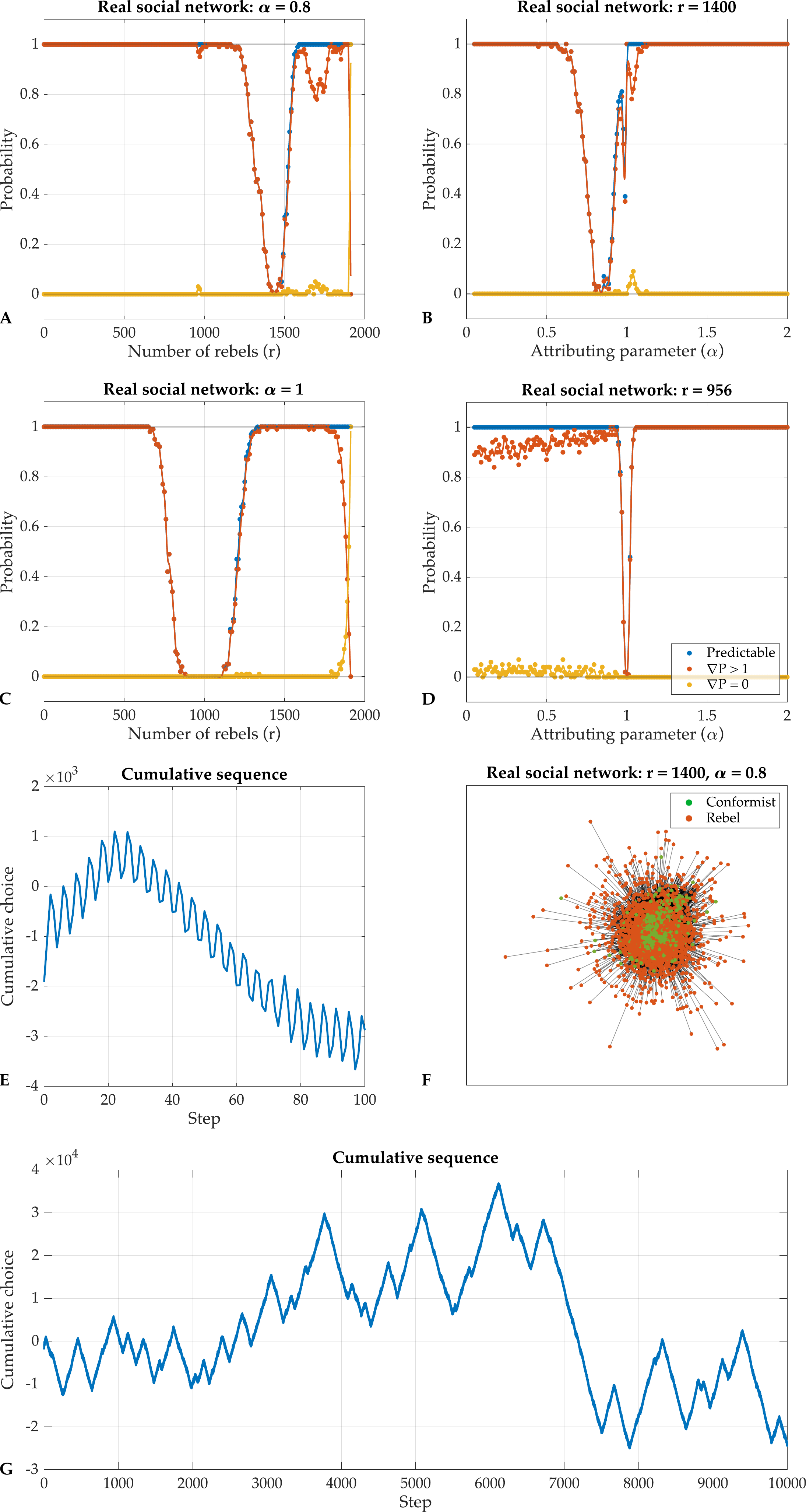}
\renewcommand{\figurename}{Figure}
\caption{{\bf Effects of the trait distribution parameters on the probability of predictable sequences with a real social network.} The relations between the probability of predictable, escalating and oscillating sequences and the number of rebels $r$ with $\alpha = 0.8$ (panel A), the attributing parameter $\alpha$ with $r = 1400$ (panel B), the number of rebels $r$ with $\alpha = 1$ (panel C) and the attributing parameter $\alpha$ with $r = 956$ (panel D); each data point is computed with 100 attributions on the real social network with initial choices $-1$ for all individuals; smooth fitted curves are added for visualization. The first 100 steps (panel E) and the first 10000 steps (panel G) of the cumulative sequence of the attributed real social network generated with $r=1400$ and $\alpha = 0.8$ (panel F).}\label{f6}
\end{center}
\end{figure}

For the real social network, if there are half of the individuals being rebels, then the rebels being uniformly distributed on the network ($\alpha = 1$) has the lowest probability of predicable sequences; see panels C and D of Figure~\ref{f6}.
If more than half of the individuals are rebels ($r = 1400$), then the lowest probability of predictable sequences occurs when the conformists are slightly scattered ($\alpha = 0.8$); see panels A and B.
In both cases, the predictable cumulative sequences generated by the real social network are almost all escalating.
We visualize an attributed network in the second case in panel F, and its cumulative sequence in panels E and G.
With 1400 rebels, the community's stable unpredictable collective decisions are driven by small constant internal conflicts.
See supplementary material for more detail.

\section{Discussion}
\label{S:4}

To model a community of individuals with different traits interactively making decisions and study the effects of network topology and trait distribution on collective decisions of the community, we have developed random processes to generate random networks with control of their size, density and heterogeneity and to attribute traits to nodes of a network with control of proportions and the extent of mixing of individuals with different traits.
With a given attributed network and initial choices, we have developed a deterministic process where each individual makes a sequence of choices following the trait-dependent rules described in Table~\ref{t1}.
The deterministic process can be considered as a cellular automaton on a graph as discussed in \cite{MARR2009}, only the cells in our process have two different types.
The process being deterministic allows us to study the cumulative sequences representing collective decisions of a community over time in addition to asymptotic or equilibrium behaviors. 
We have proved that the cumulative sequences will eventually show repeated patterns, and determined the predictability of a cumulative sequence by the appearance of the eventual repeated pattern with a computational efficient heuristic method.
The predictable cumulative sequences either escalate to an extreme or constantly oscillating, which represent collective decisions of extremizing or internally conflicting communities, while the unpredictable sequences show stable changes without extremizing or constant internal conflicts.
We have studied how network topology parameters and trait distribution parameters listed in Table~\ref{t2} affect the probability of predictable, escalating and oscillating sequences.
We have found that smaller communities, high-density communities, communities with centralized structures, communities with uneven proportions of individuals with different traits and communities with excessively clustered rebels and conformists are prone to unstable collective decisions.

To keep the model as simple as possible, we have made many unrealistic assumptions.
Social influence should have directions and varied strength, and there would also be some self-influence in reality.
We can introduce the traits of being an influencer and being a fan to the model.
We have assumed that if there are equal numbers of neighbors who chose $-1$ and $1$ in the previous step, the individual would keep the preceding choice.
In reality, individuals with traits of being conservative incline to keep the preceding choice, while progressive individuals may want to try a different choice.
We have only focused on direct social interactions and only used the information of neighbors' choices, but not indirect interactions with individuals making decisions with respect to the indices reflecting collective decisions of a community.
For example, we can introduce individuals making decisions for utility maximization.
People do not make decisions at the same time, and we have only used the information from the last step.
In reality, individuals can make decisions based on history information.
There could also be honest and dishonest individuals who would release false information to the neighbors.
These possibilities show the potential of our model as a framework for analyzing how individuals of different traits directly and indirectly interact in decision making.
It would be interesting to analyze how the proportions of individuals of different traits in a community affect the cumulative sequences and what is the role of network topology in the process. 
It would also be interesting to investigate how much of the fluctuation in, for example, stock prices, is caused by this effect of conformity and anticonformity.

In addition, the values of parameters that we chose were limited. 
Networks mainly have 100 individuals, and the random network generator can not generate all possible networks.
We defined the predictable cumulative sequences with $\nabla P > 1$ to be escalating, which can be adjusted for different standards.
It seems a limitation that cumulative sequences eventually show repeated patterns.
But people move, connections form and break, and the network topology will not stay the same forever; individuals change personalities over time and have different traits for different matters, and the trait distribution will not be unchanged either.
Therefore, the cumulative sequence of a large ever-changing community can keep being unpredictable and never show repeated patterns.

%



\section*{Implementation}
Code and data for simulations and analyses conducted in this paper are available at
\url{https://github.com/pliumath/social-interaction}

\section*{Acknowledgments}
We would like to thank Jingzhou Na for helpful discussion. 

\bibliographystyle{abbrvnat}
\bibliography{bibliography.bib}

\clearpage

\section*{Supplementary material}
 
\setcounter{figure}{0}
\setcounter{table}{0}

\subsection*{Toric lattices and random networks}
The toric lattices are boundaryless.
A node on a boundary of a lattice is related to some nodes on the opposite boundary.
Specifically, for the toric lattice of size $m$, the node at $(x,y)$ is related to the eight surrounding nodes: the eastern node at $((x+1)\text{ mod }m,y)$, the northern node at $(x,(y+1)\text{ mod }m)$, the western node at $((x-1)\text{ mod }m,y)$, the southern node at $(x,(y-1)\text{ mod }m)$, the northeastern node at $((x+1)\text{ mod }m,(y+1)\text{ mod }m)$, the northwestern node at $((x-1)\text{ mod }m,(y+1)\text{ mod }m)$, the southwestern node at $((x-1)\text{ mod }m,(y-1)\text{ mod }m)$, and the southeastern node at $((x+1)\text{ mod }m,(y-1)\text{ mod }m)$. See Supplementary Figure~\ref{sf0} for an example.

\begin{figure}[ht]
\begin{center}
\includegraphics[scale=0.5]{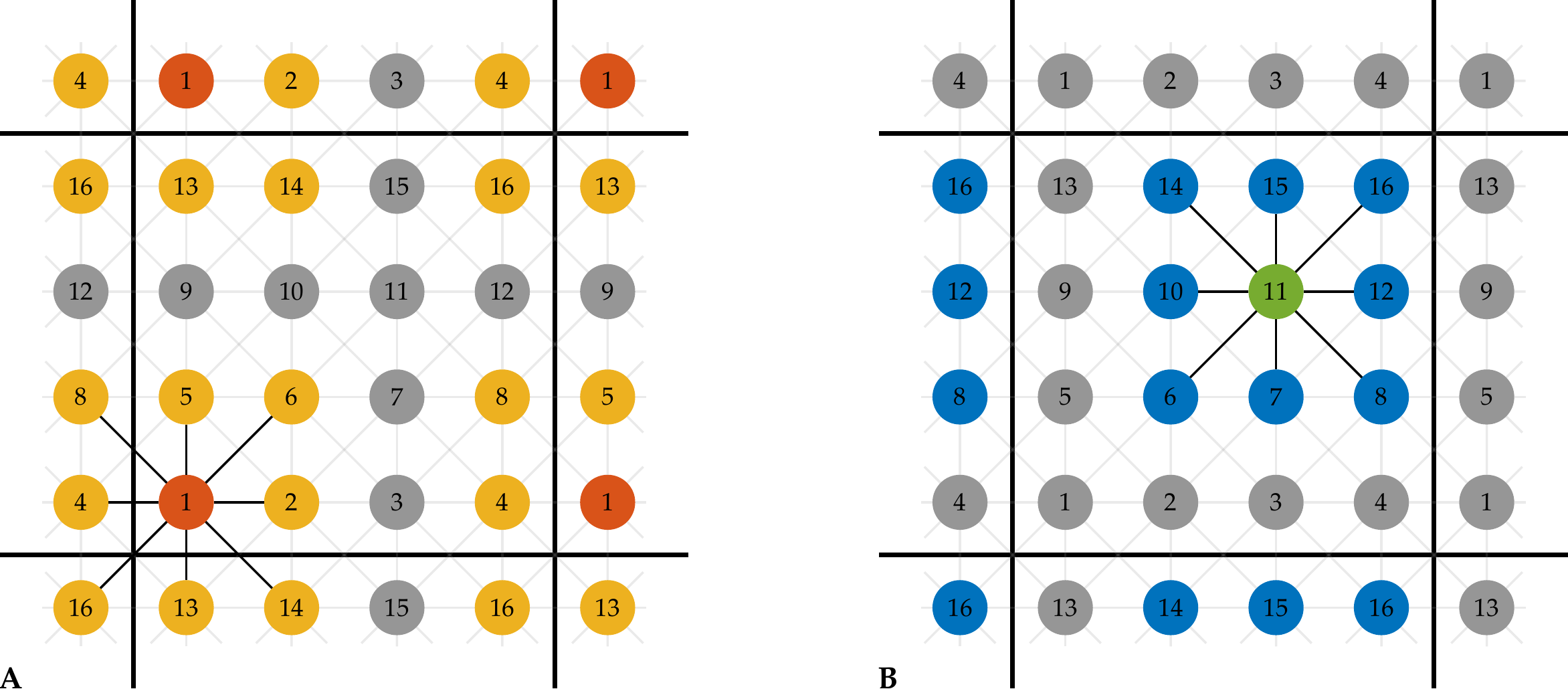}
\renewcommand{\figurename}{Supplementary Figure}
\caption{{\bf The toric lattice of size 4 and the neighbors of two individuals.} The 8 neighbors of individual 1 displayed in red are individuals 2, 4, 5, 6, 8, 13, 14, and 16 displayed in yellow (panel A). The 8 neighbors of individual 11 displayed in green are individuals 6, 7, 8, 10, 12, 14, 15, and 16 displayed in blue (panel B). }\label{sf0}
\end{center}
\end{figure}

The model generating random networks has three parameters: the number of nodes $n$, the mean degree $\mu$ and the heterogeneity parameter $\eta$ regulating the degree deviation $\sigma$. 
In Supplementary Figure~\ref{sf1}, we show the relations between the heterogeneity parameter and the degree deviation.
We generate random networks of 100 nodes with mean degree $\mu = 4$, $\mu = 8$ and $\mu = 12$ and $\eta$ ranging from $-100$ to $100$.
Each data point in Supplementary Figure~\ref{sf1} represents the average degree deviation over 1000 random networks generated with corresponding parameters, and the variance for each data point is smaller than $0.1$.

\begin{figure}[ht]
\begin{center}
\includegraphics[scale=0.5]{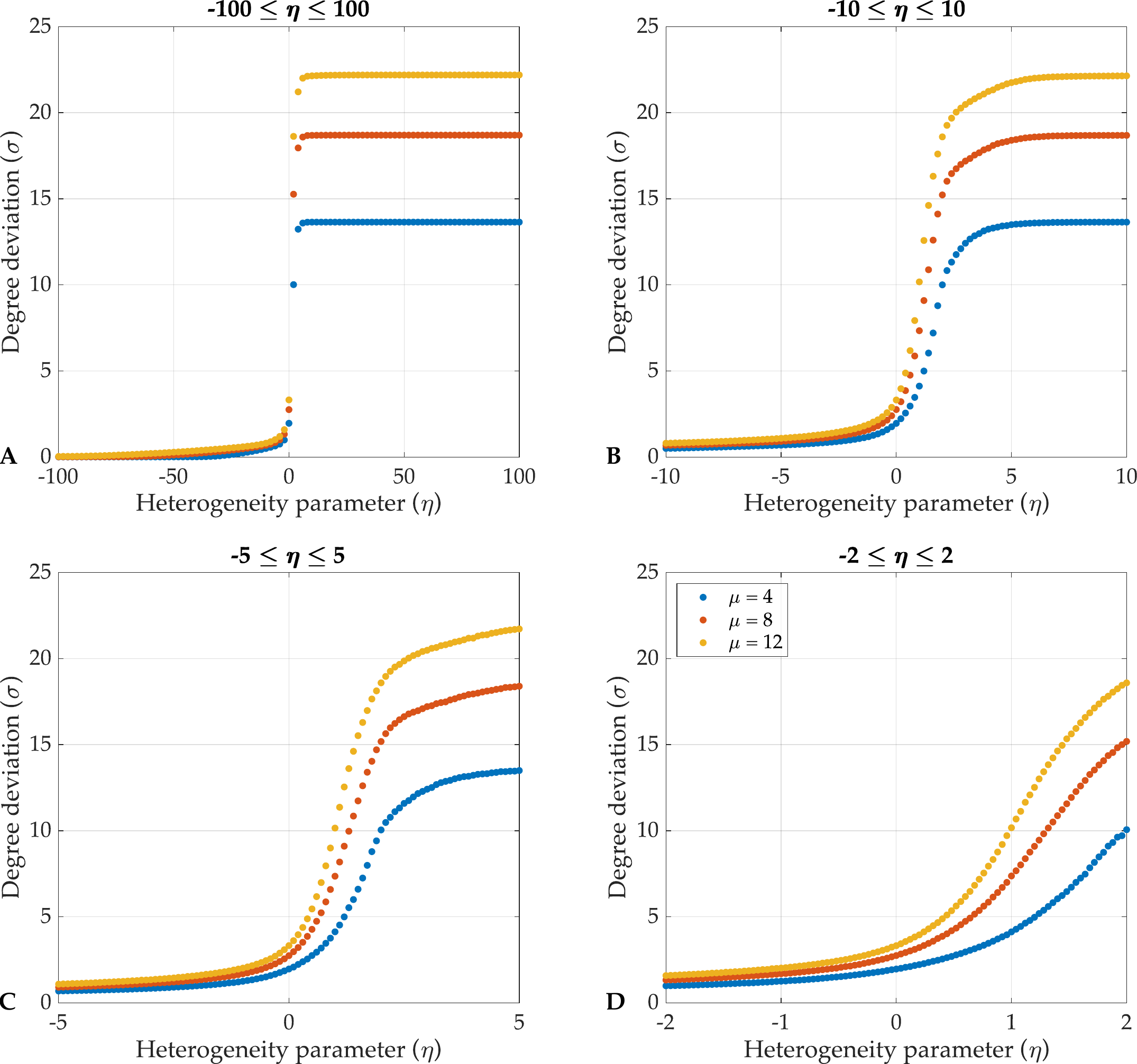}
\renewcommand{\figurename}{Supplementary Figure}
\caption{{\bf Relations between the heterogeneity parameter and degree deviations of generated random networks.} Average degree deviations of random networks of 100 nodes generated with mean degree $\mu = 4$, $\mu = 8$ and $\mu = 12$, and the heterogeneity parameter $\eta$ ranging from $-100$ to $100$ in increments of $2$ (panel A), from $-10$ to $10$ in increments of $0.2$ (panel B), from $-5$ to $5$ in increments of $0.1$ (panel C), and from $-2$ to $2$ in increments of $0.04$ (panel D). Each data point represents the average degree deviation of 1000 random networks generated with corresponding parameters. The variance for each data point is smaller than $0.1$.}\label{sf1}
\end{center}
\end{figure}

\subsection*{Trait distribution}

\begin{figure}[p]
\begin{center}
\includegraphics[scale=0.5]{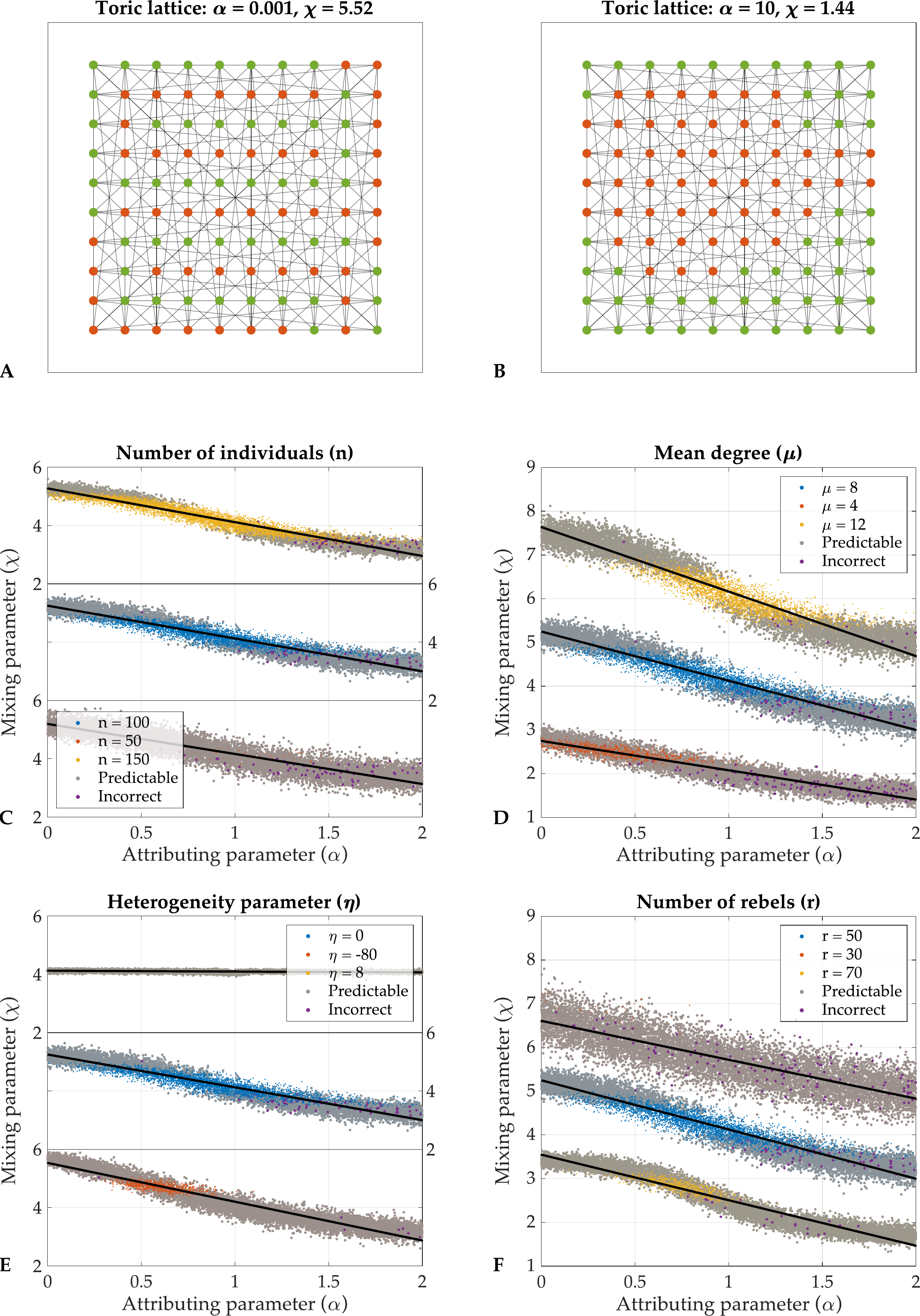}
\renewcommand{\figurename}{Supplementary Figure}
\caption{{\bf Relations between the attributing parameter and the mixing parameter of attributed networks.} Attributed toric lattices generated with scattered rebels (panel A) and clustered rebels (panel B). Relations between the attributing parameter and the mixing parameter of random networks generated with different numbers of individuals (panel C), different mean degrees (panel D), different heterogeneity parameters (panel E) and different number of rebels (panel F). Each data point represents a random network generated with corresponding parameters. For each set of parameters, 10000 random networks (data points) are generated. Grey data points represent random networks with predictable cumulative sequences, and colored data points represent random networks with unpredictable cumulative sequences. In particular, the purple data points represent the networks with unpredictable cumulative sequences that are determined to be predictable by the heuristic method.}\label{sf2}
\end{center}
\end{figure}

The process attributing traits to a network topology with the attributing parameter $\alpha$, which regulates the mixing parameter $\chi$ defined to be the average number of conformist neighbors over all rebels. 
Supplementary Figure~\ref{sf2} shows the relations between the attributing parameter and the mixing parameter of random networks. 
Panel A shows an attributed toric lattice of size $10$ generated with half of the nodes being rebels ($r = 50$) and the attributing parameter $\alpha = 0.001$. The attributed toric lattice has scattered rebels with the mixing parameter $\chi = 5.52$, which means on average a rebel has 5.52 conformist neighbors. 
Panel B shows an attributed toric lattice of size $10$ generated with half of the nodes being rebels ($r = 50$) and the attributing parameter $\alpha = 10$. The attributed toric lattice has clustered rebels with the mixing parameter $\chi = 1.44$.
Panel C displays the relations between the attributing parameter and the mixing parameter for random networks generated with different numbers of individuals $n = 50$, $n = 100$ and $n = 150$. Each data point represents an attributed random network generated with other parameters set to $\mu = 8$, $\eta = 0$, $r = 50\%n$ and the initial choices being $-1$ for all individuals.
The relations are linear in general.
For $n = 100$, we have a fitted curve $y = -1.13x + 5.25$ with $R^2 = 0.91$; for $n = 50$, we have a fitted curve $y = -1.03x + 5.19$ with $R^2 = 0.84$; for $n= 150$, we have a fitted curve $y = -1.16x + 5.27$ with $R^2 = 0.94$.
Panel D displays the relations between the attributing parameter and the mixing parameter for random networks generated with different mean degrees $\mu = 4$, $\mu = 8$ and $\mu = 12$. Each data point represents an attributed random network generated with other parameters set to $n = 100$, $\eta = 0$, $r = 50$ and the initial choices being $-1$ for all individuals.
For $\mu = 8$, we have a fitted curve $y = -1.13x + 5.25$ with $R^2 = 0.91$; for $\mu = 4$, we have a fitted curve $y = -0.67x + 2.75$ with $R^2 = 0.88$; for $\mu = 12$, we have a fitted curve $y = -1.48x + 7.64$ with $R^2 = 0.92$.
Panel E displays the relations between the attributing parameter and the mixing parameter for random networks generated with different heterogeneity parameters $\eta = -80$, $\eta = 0$ and $\eta = 8$. Each data point represents an attributed random network generated with other parameters set to $n = 100$, $\mu = 8$, $r = 50$ and the initial choices being $-1$ for all individuals.
For $\eta = 0$, we have a fitted curve $y = -1.13x + 5.25$ with $R^2 = 0.91$; for $\eta = -80$, we have a fitted curve $y = -1.33x + 5.53$ with $R^2 = 0.93$; for $\eta = 8$, we have a fitted curve $y = -0.03x + 4.12$ with $R^2 = 0.17$.
Panel F displays the relations between the attributing parameter and the mixing parameter for random networks generated with different numbers of rebels $r = 30$, $r = 50$ and $r = 70$. Each data point represents an attributed random network generated with other parameters set to $n = 100$, $\mu = 8$, $\eta = 0$ and the initial choices being $-1$ for all individuals.
For $r = 50$, we have a fitted curve $y = -1.13x + 5.25$ with $R^2 = 0.91$; for $r = 30$, we have a fitted curve $y = -0.89x + 6.61$ with $R^2 = 0.72$; for $r = 70$, we have a fitted curve $y = -1.04x + 3.55$ with $R^2 = 0.94$.

\subsection*{Predictability of cumulative sequences}

We argue that every collective sequence of choices eventually enters a unique period.
For a community of $n$ individuals, there are $2^n$ unique choice patterns. 
Note that the deterministic process defined in Section~\ref{S:2.1} is memoryless in the sense that the choice pattern at step $k$ only depends on the choice pattern at step $k-1$.
Moreover, each choice pattern determines a unique succeeding choice pattern.
If the deterministic process has more than $2^n$ steps, then the sequence of choice patterns must have identical elements $C(\cdot,k) = C(\cdot,l)$ due to the pigeon hole principle. 
Since the process is deterministic, identical subsequences of choice patterns follow $C(\cdot,k)$ and $C(\cdot,l)$, and periodicity appears in the sequence of choice patterns, hence in the collective sequence of the community's choices.
Thus, given the network topology, the trait distribution and initial choices for a community of individuals, the collective sequence eventually enters a unique period determined by the three factors. 

Recall that the length $L(P)$ of the eventual period $P$ and the length $L(Q)$ of the pre-period subsequence $Q$ of a cumulative sequence are defined to be the numbers of steps that $P$ and $Q$ span respectively.
The period gain $\Delta P$ of the eventual period is the change in cumulative sequence over the period $P$.
The gradient of the eventual period is defined to be $\nabla P = \abs{\Delta P}/L(P)$.
In Supplementary Figure~\ref{sf3}, we show the eventual periods and the pre-period subsequences of cumulative sequences of choices of two communities.
Panel A shows the first 10000 steps of the cumulative sequence of choices of the community represented by the attributed toric lattice displayed in panel B.
The attributed toric lattice has size $m = 10$, and the trait distribution is generated with $r = 50$ and $\alpha = 0.37$.
The initial choices of the community are $-1$ for all individuals.
In panel A, the first 10000 steps of the cumulative sequence do not contain a complete eventual period, so the cumulative sequence is unpredictable. 
Actually, the pre-period subsequence $Q$ displayed in panel C before the first vertical line has length $L(Q) = 11010$.
If we extend the length of the process to $t = 160000$, we see three complete eventual periods of the cumulative sequence displayed in panel C.
The eventual period $P$ showed in panel C has length $L(P)=47115$, period gain $\Delta P = -474$ and gradient $\nabla P = 0.01$. 
Similarly, The attributed toric lattice displayed in panel E has size $m = 10$, and the trait distribution is generated with $r = 50$ and $\alpha = 0.7$.
The initial choices of the community are $-1$ for all individuals.
Panel D and panel F show the pre-period subsequence $Q$ of length $L(Q) = 17749$, and the eventual period $P$ has length $L(P)=1$, period gain $\Delta P = -8$ and gradient $\nabla P = 8$. 

\begin{figure}[p]
\begin{center}
\includegraphics[scale=0.5]{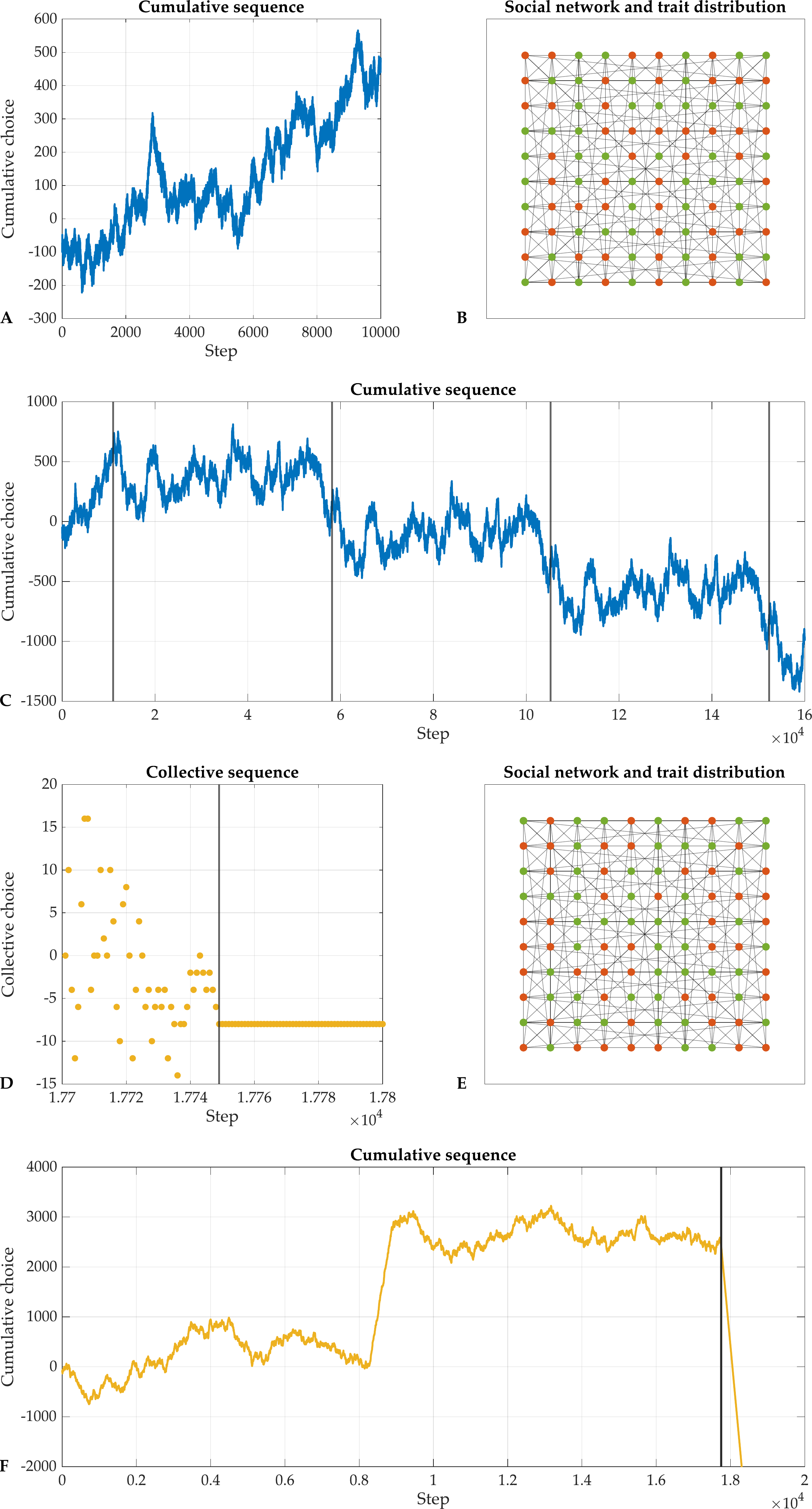}
\renewcommand{\figurename}{Supplementary Figure}
\caption{{\bf Eventual periods and pre-period subsequences of cumulative sequences.} The first 10000 steps of the cumulative sequence (panel A) of the attributed network generated with $m = 10$, $r = 50$ and $\alpha = 0.37$ (panel B) and the first 160000 steps of the cumulative sequence (panel C).
The 100 steps of collective sequence (panel D) near the appearance of the first eventual period for the attributed network generated with $m = 10$, $r = 50$ and $\alpha = 0.7$ (panel E) and the first 20000 steps of the cumulative sequence (panel F).
The complete eventual periods are displayed between vertical lines, and the pre-period subsequences are before the first vertical line in panels C and F.
The green nodes represent conformists and the red nodes represent rebels in panels B and D.}\label{sf3}
\end{center}
\end{figure}

To efficiently determine if a collective sequence $s = [s(0),s(1),...,s(t)]$ is predictable without recording and comparing choice patterns, we develop the heuristic method as follows.
We extract the subsequence $s' = [s(t+1-\tau),...,s(t-1),s(t)]$ consisting of the last $\tau$ elements in $s$ and search for subsequences of $s$ with $\tau$ consecutive elements that are identical to $s'$.
If $s'$ is the only subsequence, then the heuristic method determines the collective sequence and the corresponding cumulative sequence to be unpredictable.
If there are more than one subsequences in $s$ that are identical to $s'$, then the heuristic method determines the collective sequence and the corresponding cumulative sequence to be predictable.

We argue that the heuristic method faithfully determines every predictable collective sequence.
Let $s$ be a predictable collective sequence and $s'$ be the subsequence of $s$ consisting of the last $\tau$ elements.
By definition, there exist at least one complete eventual period in the first $t+1-\tau$ steps of $s$. 
If there exists one eventual period in the first $t+1-\tau$ steps, then what follows must be in the eventual period. 
Hence, $s'$ must be in the eventual period, and there exists at least one subsequence in the first $t+1-\tau$ steps that is identical to $s'$.
Therefore, the heuristic method faithfully determines $s$ to be predictable.
If $s$ is unpredictable, that is, there exists no eventual period in the first $t+1-\tau$ steps, then there may still be subsequences in the first $t+1-\tau$ steps that are identical to $s'$. 
This is because different choice patterns may have the same sum of choices. 
So, the heuristic method may incorrectly determine $s$ to be predictable and underestimate the probability of cumulative sequences being unpredictable.

In Supplementary Figure~\ref{sf2}, we show the unpredictable collective sequences that are incorrectly determined by the heuristic method as predictable ones with purple data points. 
In panel C, there are 75 data points with incorrect predictability for $n=100$, 70 for $n=50$ and 65 for $n=150$
In panel D, there are 124 data points with incorrect predictability for $\mu = 4$ and 21 for $\mu = 12$.
In panel E, there are 8 data points with incorrect predictability for $\eta = -80$ and 0 for $\eta = 8$.
In panel F, there are 136 data points with incorrect predictability for $r = 30$ and 22 for $r = 70$.
On average, $0.58\%$ of the unpredictable collective sequences are incorrectly determined to be predictable by the heuristic method.

\subsection*{Homogeneous attributed networks}

We can deduce the cumulative sequences for the four homogeneous attributed networks displayed in Supplementary Figure~\ref{sf4}.
The toric lattice of size $m = 10$ displayed in panel B is attributed with $r = 50$, and the conformists and rebels are homogeneously separated into two clusters.
Each rebel in the interior of the cluster has 8 rebel neighbors, and each rebel on the boundary of the cluster has 5 rebel neighbors and 3 conformist neighbors. 
The cluster of conformists also have the same patterns.
When the initial choices are $-1$ for all individuals, the two clusters can not affect each other, so all rebels will change at every step, and all conformists will keep choosing $-1$ at every step.
Therefore, the attributed network generates an escalating cumulative sequence displayed in panel A with $\nabla P = 50$, as the collective sequence consists of alternating $0$ and $-100$.
The toric lattice of size $m = 10$ displayed in panel D is attributed with $r = 50$, and the individuals with different traits are homogeneously mixed such that every rebel has 6 conformist neighbors and 2 rebel neighbors, and symmetrically, every conformist has 6 rebel neighbors and 2 conformist neighbors.
Suppose that the initial choices are $-1$ for all individuals.
At step 1, the rebels will choose $1$, and the conformists will keep $-1$; at step 2, the rebels will keep $1$, and the conformists will choose $1$; at step 3, the rebels will choose $-1$, and the conformists will keep $1$; at step 4, the rebels will keep $-1$, and the conformists will choose $-1$, which is the same choice pattern as the initial choices. 
Thus, the attributed network generates an oscillating cumulative sequence displayed in panel C with $\nabla P = 0$.
The connected random network displayed in panel F has 100 conformist individuals.
When the initial choices are $-1$ for all individuals, all conformists will keep choosing $-1$ at every step, and the attributed network generates an escalating cumulative sequence displayed in panel E with $\nabla P = 100$.
The connected random network displayed in panel H has 100 rebel individuals.
When the initial choices are $-1$ for all individuals, all rebels will change their choices at every step, and the attributed network generates an oscillating cumulative sequence displayed in panel G with $\nabla P = 0$.

\begin{figure}[h!p]
\begin{center}
\includegraphics[scale=0.5]{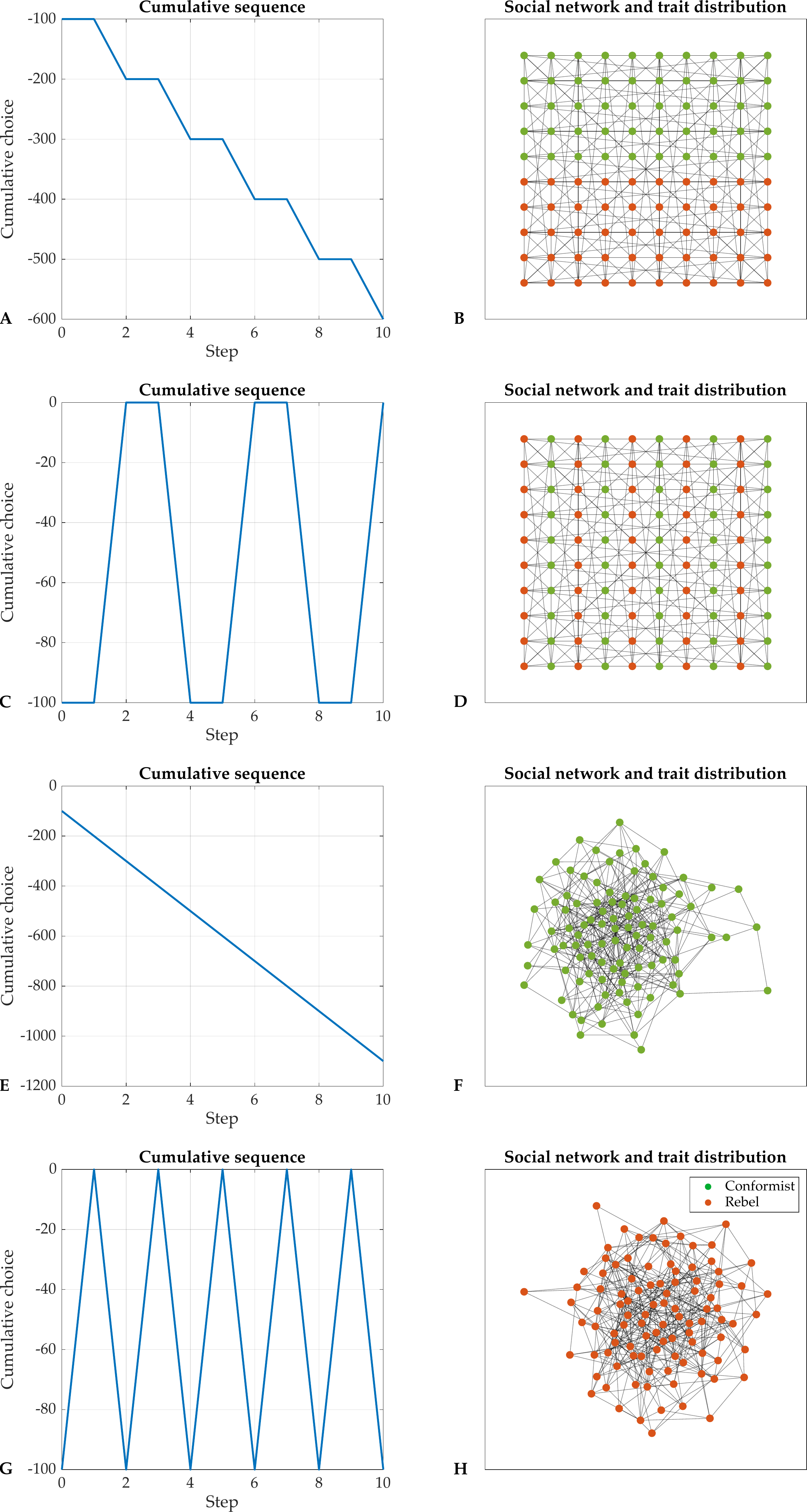}
\renewcommand{\figurename}{Supplementary Figure}
\caption{{\bf Homogeneous attributed networks and the corresponding cumulative sequences.} The escalating cumulative sequence (panel A) of the toric lattice with homogeneously clustered conformists and rebels (panel B). The oscillating cumulative sequence (panel C) of the toric lattice with homogeneously mixed conformists and rebels (panel D). The escalating cumulative sequence (panel E) of the random network of all conformists (panel F). The oscillating cumulative sequence (panel G) of the random network of all rebels (panel H). The initial choices for the four networks are $-1$ for all individuals. }\label{sf4}
\end{center}
\end{figure}

\subsection*{Results in alternative settings}
We investigate the effects of the three network topology parameters and the two trait distribution parameters on the probability of predictable, escalating and oscillating sequences in different settings. 
See Supplementary Figure~\ref{sf5} - \ref{sf9}.
The number of individuals has a default value $n = 100$.
In investigating effects of the other four parameters, we set $n = 50$ and $n = 150$.
The mean degree has a default value $\mu = 8$.
In investigating effects of the other four parameters, we set $\mu = 4$ and $\mu = 12$.
The heterogeneity parameter has a default value $\eta = 0$.
In investigating effects of the other four parameters, we set $\eta = -80$ and $\eta = 8$.
The number of rebels has a default value of $r = 50\%n$.
In investigating effects of the other four parameters, we set $r = 40\%n$ and $r = 70\%n$.
The attributing parameter has a default value of $\alpha = 0.8$.
In investigating effects of the other four parameters, we set $\alpha = 0.6$ and $\alpha = 1$.

We also generated random networks with parameters $n = 1912$, $\mu = 32.74$ and $\eta = 1.19$ that resembles the parameters of the real social network. 
We choose $\eta = 1.19$ so that the generated random networks have degree deviation near $\sigma = 55.85$. 
In Supplementary Figure~\ref{sf10}, we display the relations and an attributed real social network with $r = 956$ and $\alpha = 1$ and its cumulative sequence.

\begin{figure}[ht]
\begin{center}
\includegraphics[scale=0.5]{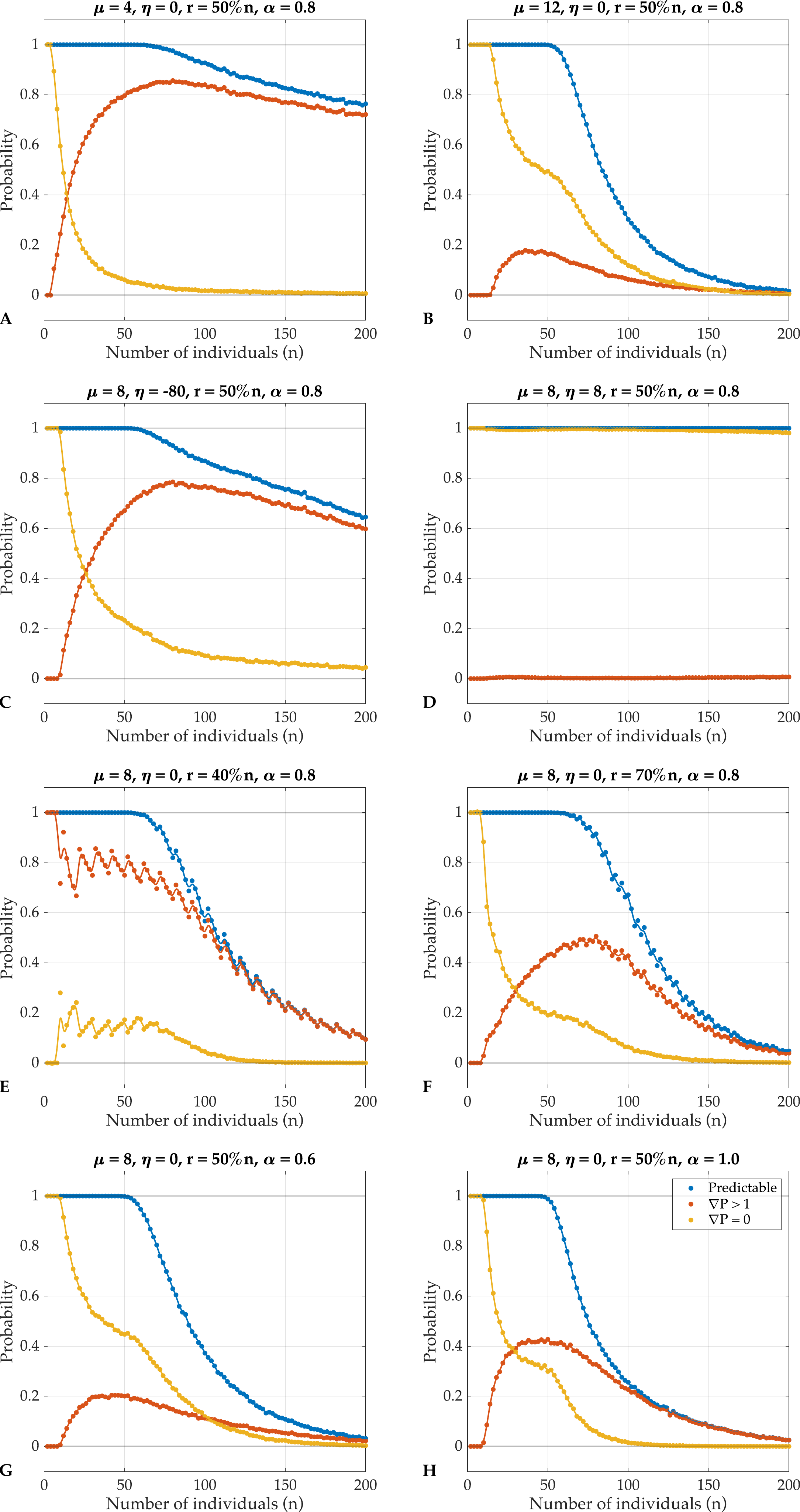}
\renewcommand{\figurename}{Supplementary Figure}
\caption{{\bf Effects of the number of individuals on the probability of predictable sequences in different settings.} The relations between the probability of predictable, escalating and oscillating sequences and the number of individuals $n$ for random networks with $\mu=4$ (panel A) and $\mu = 12$ (panel B), networks with $\eta = -80$ (panel C) and $\eta = 8$ (panel D), networks with $r = 40$ (panel E) and $r = 70$ (panel F) and networks with $\alpha = 0.6$ (panel G) and $\alpha = 1$ (panel H).  Each data point is computed with 10000 random networks and their corresponding cumulative sequences with initial choices $-1$ for all individuals. Smooth fitted curves are added for visualization. }\label{sf5}
\end{center}
\end{figure}

\begin{figure}[ht]
\begin{center}
\includegraphics[scale=0.5]{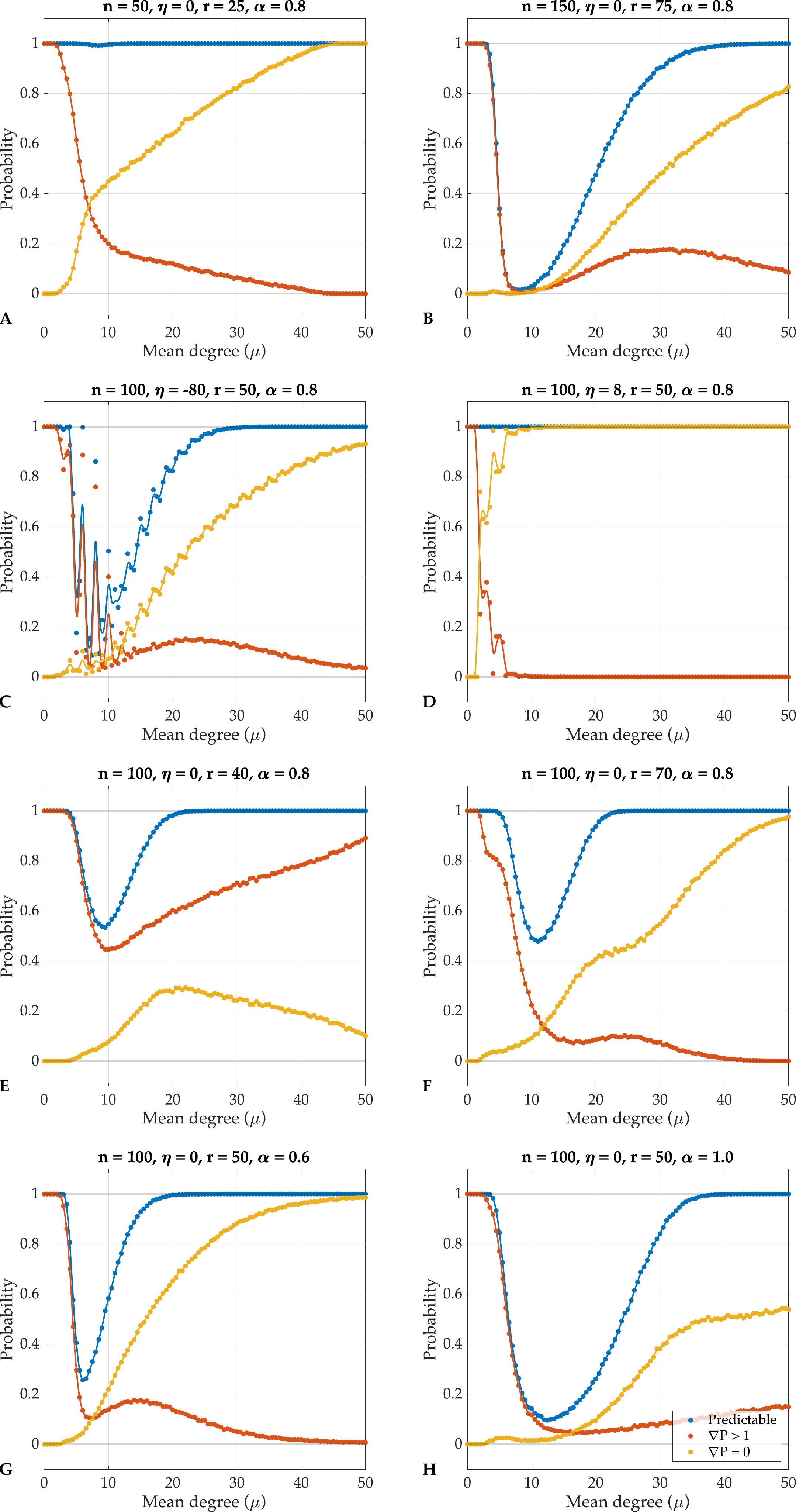}
\renewcommand{\figurename}{Supplementary Figure}
\caption{{\bf Effects of the mean degree on the probability of predictable sequences in different settings.} The relations between the probability of predictable, escalating and oscillating sequences and the mean degree $\mu$ for random networks with $n=50$ (panel A) and $n = 150$ (panel B), networks with $\eta = -80$ (panel C) and $\eta = 8$ (panel D), networks with $r = 40$ (panel E) and $r = 70$ (panel F) and networks with $\alpha = 0.6$ (panel G) and $\alpha = 1$ (panel H).  Each data point is computed with 10000 random networks and their corresponding cumulative sequences with initial choices $-1$ for all individuals. Smooth fitted curves are added for visualization.}\label{sf6}
\end{center}
\end{figure}

\begin{figure}[ht]
\begin{center}
\includegraphics[scale=0.5]{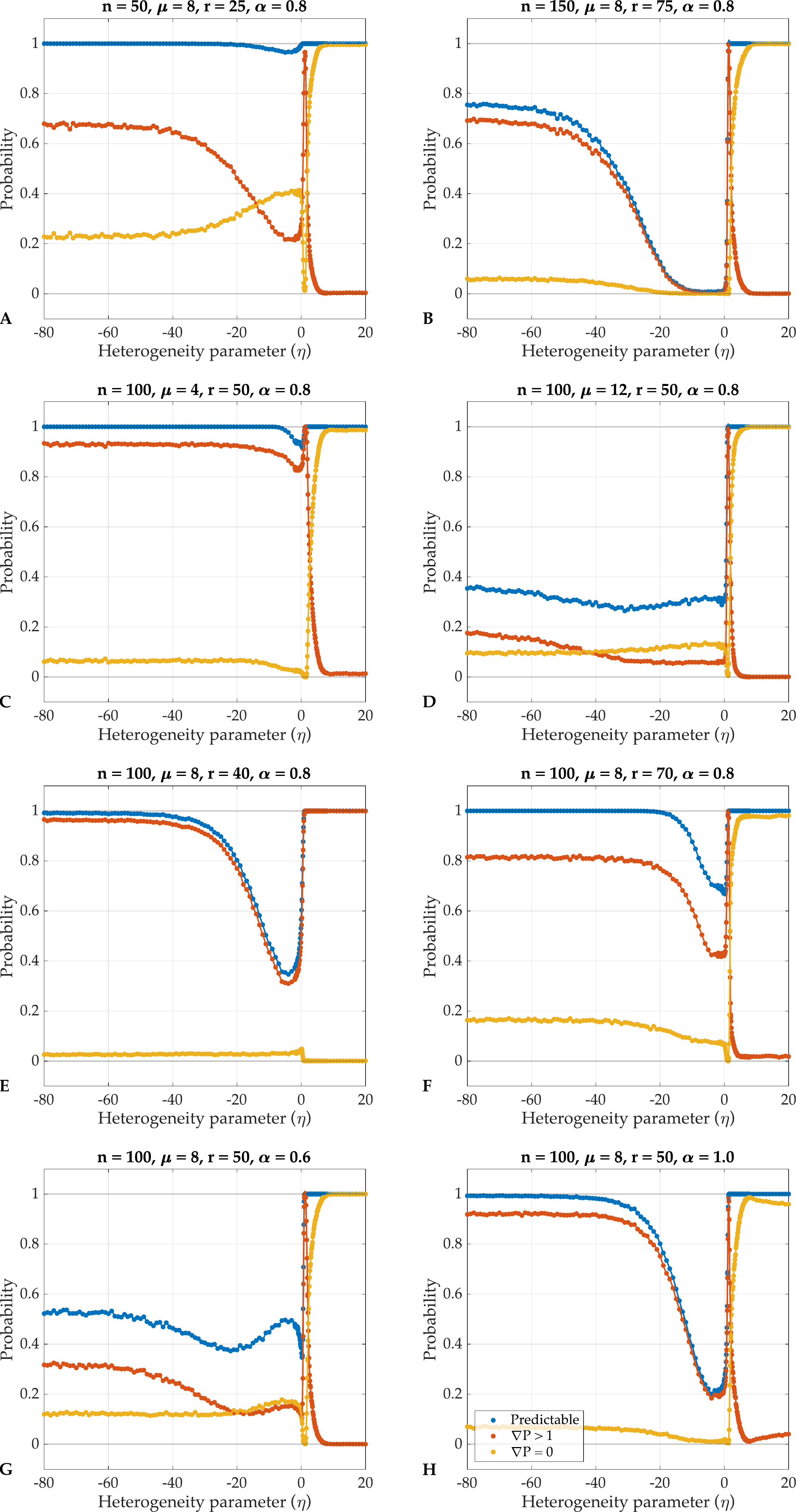}
\renewcommand{\figurename}{Supplementary Figure}
\caption{{\bf Effects of the heterogeneity parameter on the probability of predictable sequences in different settings.} The relations between the probability of predictable, escalating and oscillating sequences and the heterogeneity parameter $\eta$ for random networks with $n=50$ (panel A) and $n = 150$ (panel B), networks with $\mu =4$ (panel C) and $\mu = 12$ (panel D), networks with $r = 40$ (panel E) and $r = 70$ (panel F) and networks with $\alpha = 0.6$ (panel G) and $\alpha = 1$ (panel H).  Each data point is computed with 10000 random networks and their corresponding cumulative sequences with initial choices $-1$ for all individuals. Smooth fitted curves are added for visualization.}\label{sf7}
\end{center}
\end{figure}

\begin{figure}[ht]
\begin{center}
\includegraphics[scale=0.5]{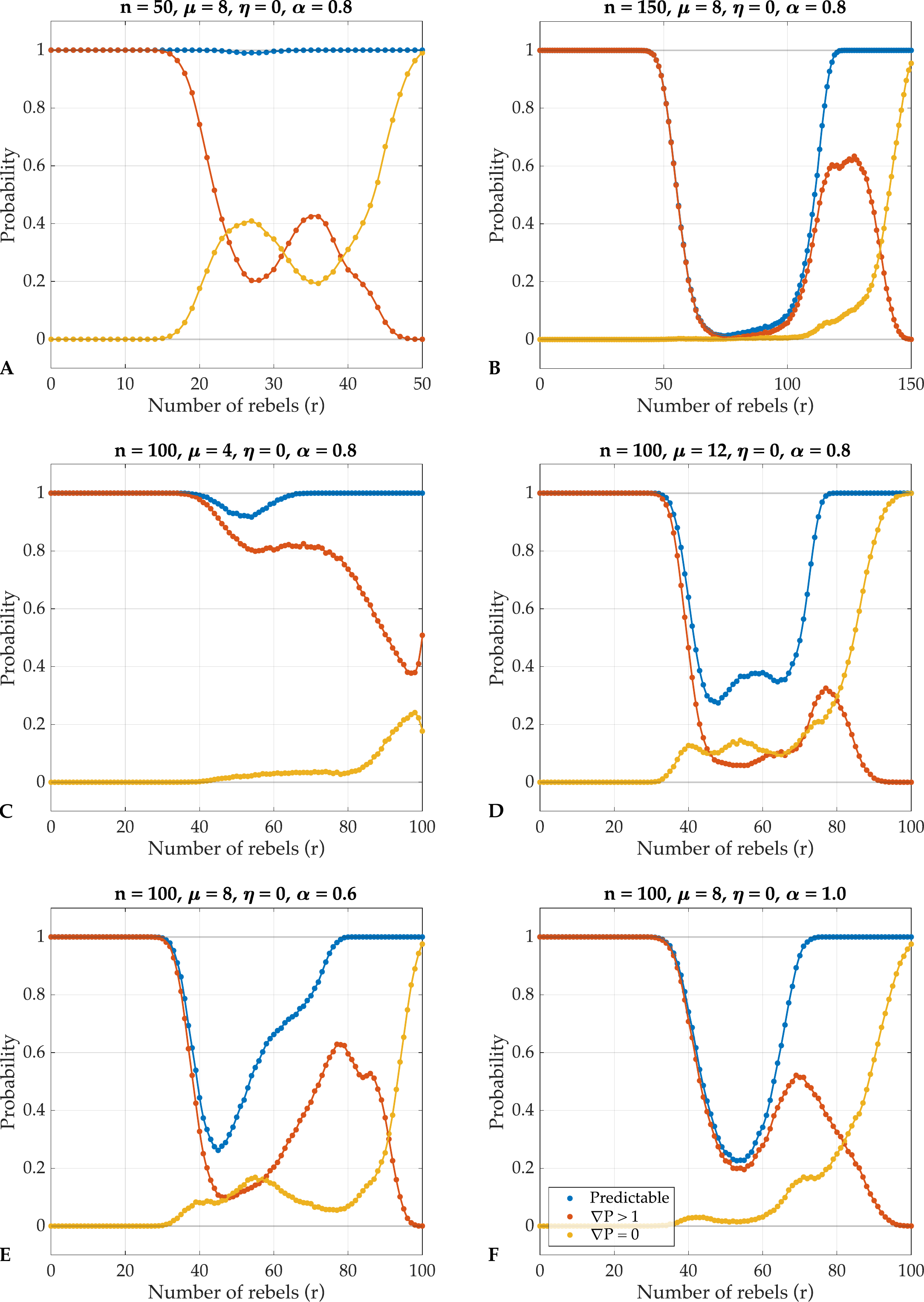}
\renewcommand{\figurename}{Supplementary Figure}
\caption{{\bf Effects of the number of rebels on the probability of predictable sequences in different settings.} The relations between the probability of predictable, escalating and oscillating sequences and the attributing parameter $\alpha$ for random networks with $n=50$ (panel A) and $n = 150$ (panel B), networks with $\mu =4$ (panel C) and $\mu = 12$ (panel D) and networks with $r = 40$ (panel E) and $r = 70$ (panel F).  Each data point is computed with 10000 random networks and their corresponding cumulative sequences with initial choices $-1$ for all individuals. Smooth fitted curves are added for visualization.}\label{sf8}
\end{center}
\end{figure}

\begin{figure}[ht]
\begin{center}
\includegraphics[scale=0.5]{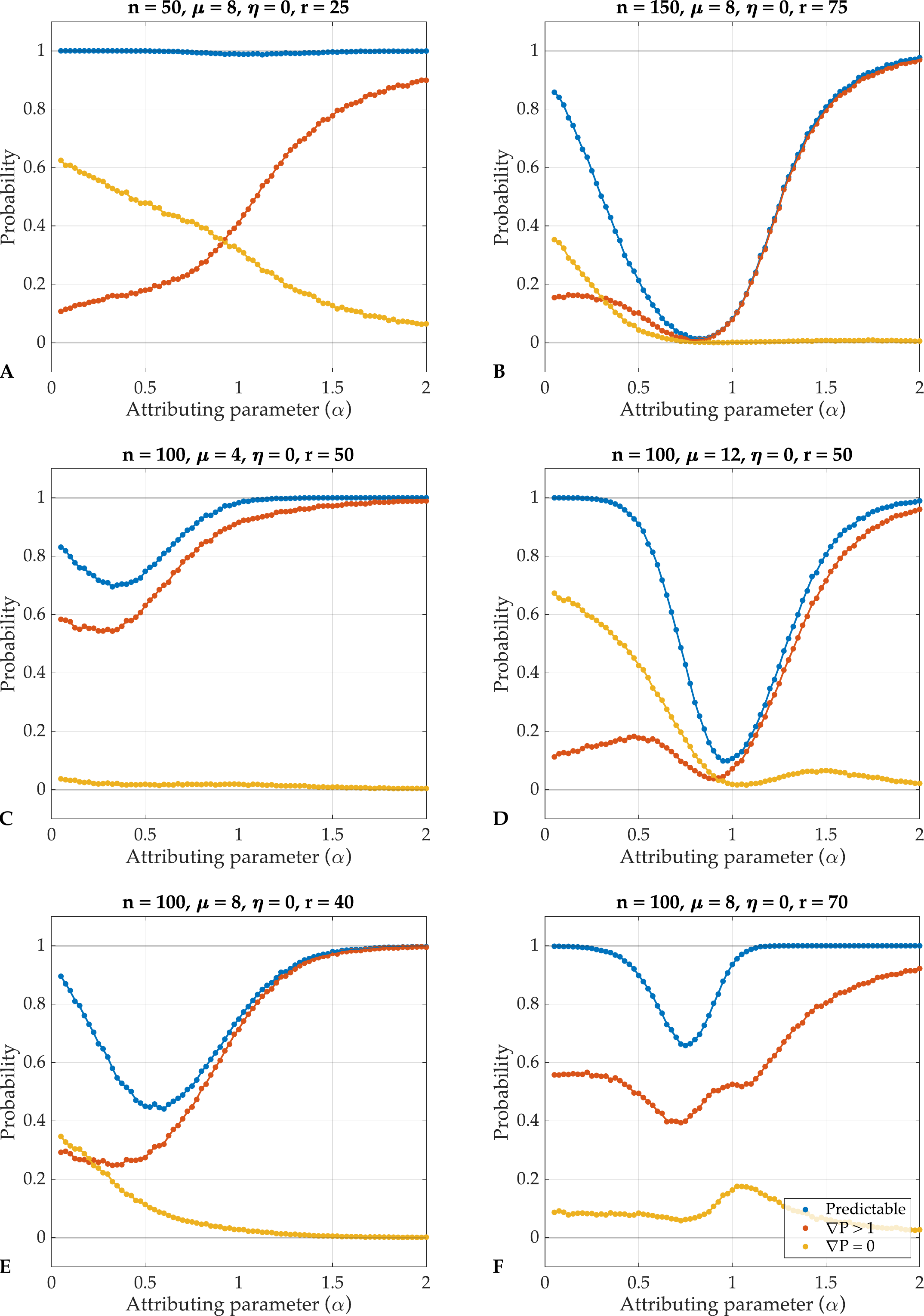}
\renewcommand{\figurename}{Supplementary Figure}
\caption{{\bf Effects of the attributing parameter on the probability of predictable sequences in different settings.} The relations between the probability of predictable, escalating and oscillating sequences and the number of rebels $r$ for random networks with $n=50$ (panel A) and $n = 150$ (panel B), networks with $\mu =4$ (panel C) and $\mu = 12$ (panel D) and networks with $\alpha = 0.6$ (panel E) and $\alpha = 1$ (panel F).  Each data point is computed with 10000 random networks and their corresponding cumulative sequences with initial choices $-1$ for all individuals. Smooth fitted curves are added for visualization.}\label{sf9}
\end{center}
\end{figure}

\begin{figure}[ht]
\begin{center}
\includegraphics[scale=0.5]{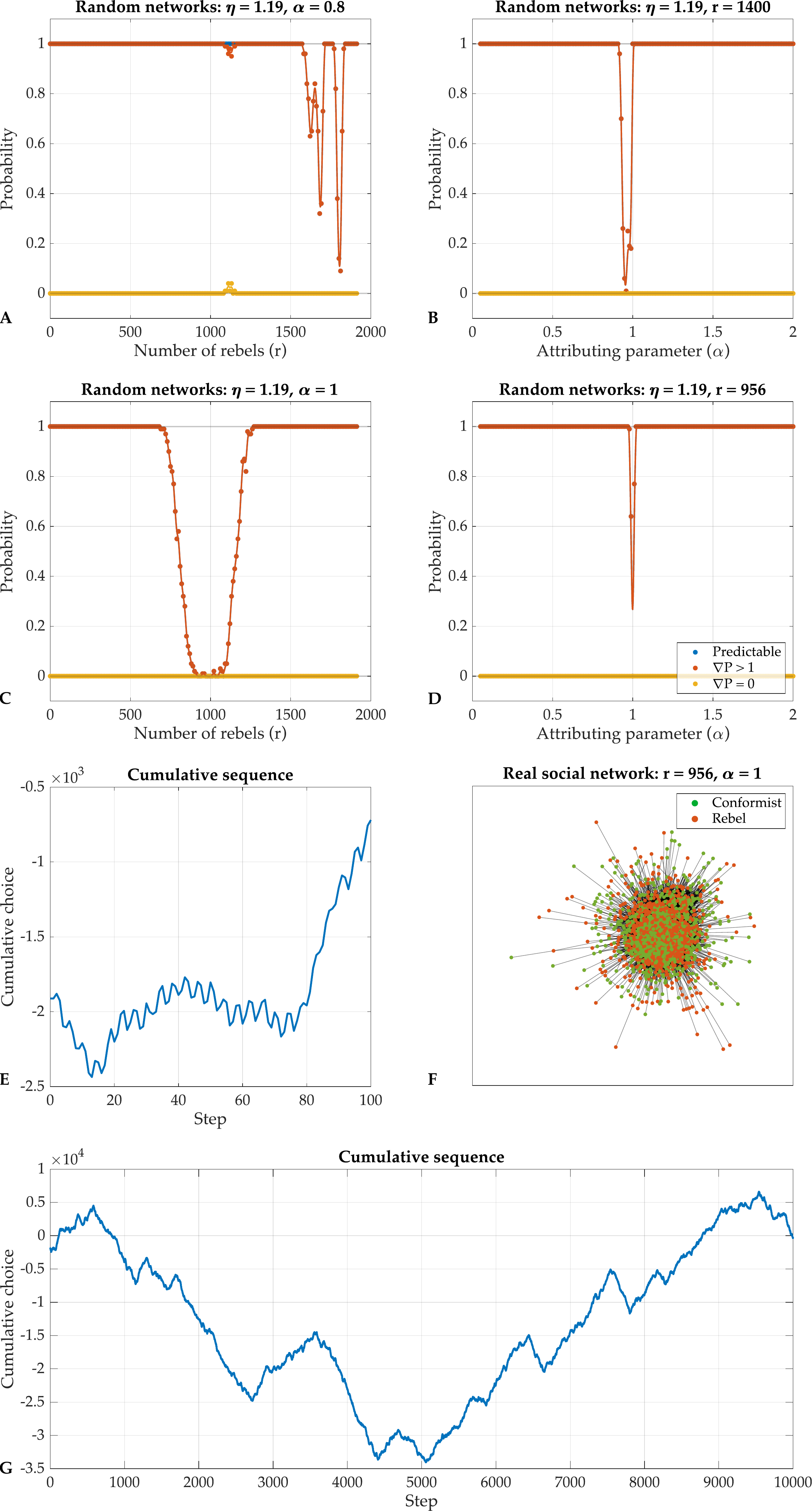}
\renewcommand{\figurename}{Supplementary Figure}
\caption{{\bf Effects on random networks generated with parameters resembling the real social network.} The relations between the probability of predictable, escalating and oscillating sequences and the number of rebels $r$ with $\alpha = 0.8$ (panel A), the attributing parameter $\alpha$ with $r = 1400$ (panel B), the number of rebels $r$ with $\alpha = 1$ (panel C) and the attributing parameter $\alpha$ with $r = 956$ (panel D) for random networks with $\eta = 1.19$; each data point is computed with 100 attributions on the real social network with initial choices $-1$ for all individuals; smooth fitted curves are added for visualization. The first 100 steps (panel E) and the first 10000 steps (panel G) of the cumulative sequence of the attributed real social network generated with $r=956$ and $\alpha = 1$ (panel F). }\label{sf10}
\end{center}
\end{figure}

\end{document}